\documentclass[prx,aps,twocolumn,amsmath,amssymb,floatfix]{revtex4-1}

\pdfoutput=1
\usepackage{hyperref}
\usepackage{graphicx}
\usepackage[usenames]{color}
\usepackage{bm}

\def\cal#1{\mathcal{#1}}

\def\eq#1{(\ref{#1})}

\def\f#1{Fig.~\ref{#1}}

\def\s#1{Section~\ref{#1}}

\def\c#1{~\cite{#1}}

\def\cc#1{Ref.~\cite{#1}}
\def\ccc#1{Refs.~\cite{#1}}

\def\x{\bm x}
\def\y{\bm y}

\def\beq{\begin{equation}}
\def\eeq{\end{equation}}
\def\bea{\begin{eqnarray}}
\def\eea{\end{eqnarray}}

\def\kt{k_{\rm B}T}

\begin{document}

\title{Neuroevolutionary learning of particles and protocols for self-assembly}
\author{Stephen Whitelam}\email{swhitelam@lbl.gov}
\affiliation{Molecular Foundry, Lawrence Berkeley National Laboratory, 1 Cyclotron Road, Berkeley, CA 94720, USA}
\author{Isaac Tamblyn}\email{isaac.tamblyn@nrc.ca}
\affiliation{National Research Council of Canada, Ottawa, ON, Canada}
\affiliation{Vector Institute for Artificial Intelligence, Toronto, ON, Canada}

\begin{abstract}
Within simulations of molecules deposited on a surface we show that neuroevolutionary learning can design particles and time-dependent protocols to promote self-assembly, without input from physical concepts such as thermal equilibrium or mechanical stability and without prior knowledge of candidate or competing structures. The learning algorithm is capable of both directed and exploratory design: it can assemble a material with a user-defined property, or search for novelty in the space of specified order parameters. In the latter mode it explores the space of what can be made rather than the space of structures that are low in energy but not necessarily kinetically accessible.
\end{abstract}
\maketitle

{\em Introduction---} How do we make a material with specified properties? In pursuit of ``synthesis by design''\c{de2016basic,broholm2016basic} the materials science community has developed and adapted algorithms of inverse design and machine learning. These approaches can identify interparticle potentials able to stabilize target structures or promote their self-assembly from solution\c{cohn2009algorithmic,torquato2009inverse,bianchi2012predicting,rechtsman2006self,lindquist2016communication,jadrich2017probabilistic,long2018rational,gadelrab2017inverting,ferguson2017machine,pineros2018inverse,van2015digital,adorf2018inverse,madge2017optimising,jiang2018evolutionary,kumar2019inverse,zhou2019alchemical,sherman2020inverse,reinhardt2014numerical,romano2020designing,miskin2016turning}, and can identify protocols or reaction conditions that optimize the self-assembly of specified particles\c{klotsa2013controlling,miskin2016turning,raccuglia2016machine,whitelam2020learning,tang2016optimal}. 

Here we present an approach based on evolutionary learning\c{GA} that simultaneously designs particles {\em and} protocols in order to self-assemble materials to order. We study a coarse-grained computational model of molecular self-assembly at a surface\c{bartels2010tailoring,elemans2009molecular,swiegers2002classification}. Coarse-grained models are simple by design\c{doye2004inhibition,hagan2006dynamic,molinero2008water,romano2011colloidal,glotzer2004self,doye2007condensed,rapaport2010modeling,murugan2015undesired,whitelam2015statistical,grunwald2014patterns,nguyen2016design,lutsko2019crystals,fan2019orientational,carpenter2020heterogeneous} but can exhibit key features of real systems, including the formation of complex structures and kinetic traps that impair assembly\c{thorkelsson2015self,biancaniello2005colloidal,park2008dna,nykypanchuk2008dna,pfeifer2018synthetic,de2015crystallization}. The particular class of model we use here has been shown to reproduce the thermodynamic and dynamic behavior of a range of molecular and nanoscale assemblies at surfaces\c{whitelam2014common}. Such models provide a rigorous test of algorithmic control of self-assembly. 

In order to allow thorough exploration of the self-assembly behavior accessible to this class of model we express the interparticle potential and time-dependent assembly protocol as arbitrary functions, encoded by neural networks. In evolutionary language, which reflects the method of learning used and provides a mnemonic for the role of each component of the algorithm, this encoding is the instruction code or ``genome'' for self-assembling a material. Molecular simulation carried out using the particle and protocols specified by the genome results in the ``phenome'', a material whose properties can be measured and compared to a design goal. 

Evolutionary learning on the parameters of the neural networks, called neuroevolution\c{GA,GA2,salimans2017evolution,montana1989training,Guber,whitelam2020correspondence}, can produce materials whose properties satisfy user-defined goals, which can be directed or exploratory in nature. The algorithm needs no information about possible candidate or competing structures, nor prior knowledge of what constitutes a good self-assembly protocol or particle design. For directed design we specify materials with certain pore geometries and isolated clusters of certain sizes. The solutions identified by the learning algorithm include interparticle interactions whose symmetries can be realized by known molecules or complexes\c{swiegers2002classification,bartels2010tailoring,elemans2009molecular}. The approach is simple to implement and can identify sophisticated design strategies, realizing complex structures via hierarchical self-assembly pathways. It can also fail, if the design goal is too challenging, and we discuss how to modify the goal in such cases.
\begin{figure*}[] 
   \centering
\includegraphics[width=\linewidth]{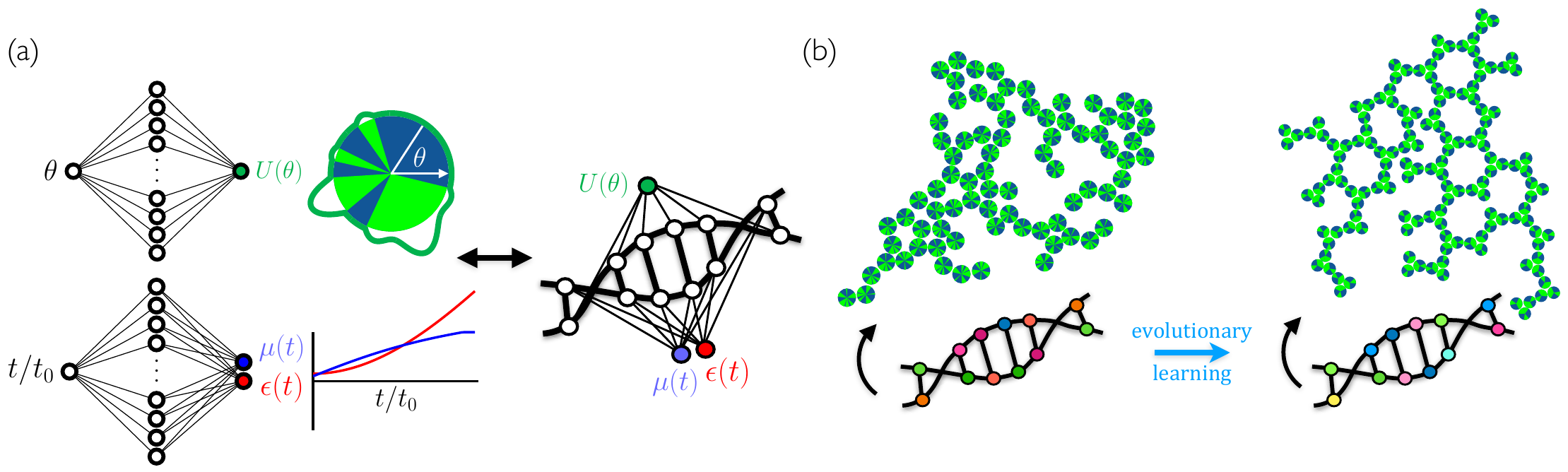} 
   \caption{\label{fig_schematic} (a) We express the angular interaction potential and the time-dependent self-assembly protocol for a set of model molecules in the form of two neural networks, which together comprise the ``genome'' for making a material. In images, attractive portions of particles are shown green. (b) Upon specifying a design goal, neuroevolutionary learning can produce a genome whose ``phenome'' -- the result of molecular simulations carried out using the potential and protocol specified by the genome -- is a material satisfying that goal.}
\end{figure*}

Our approach is similar to that of \ccc{lindquist2016communication,jadrich2017probabilistic} in that we use an iterative learning method to promote self-assembly, but differs in that we do not specify or build the target structure in advance. In that respect it is similar to the approach of~\cc{miskin2016turning}, and complementary to that work in that we express the design problem differently (in the form of neural networks) and optimize differently (via evolutionary methods). Our approach differs from other approaches to inverse design in that we consider the design of particles and protocols simultaneously, and do not appeal to physical concepts such as thermal equilibrium or mechanical stability. We also go beyond traditional forms of inverse design and borrow from the machine-learning literature to specify the design goal of {\em novelty}\c{conti2018improving}. Within a space specified by certain order parameters we instruct the evolutionary learning algorithm to produce materials not seen previously, rather than materials with particular properties. If we specify materials with 3- and 4-membered pores then novelty search identifies structures dual to regular and semi-regular tilings of the plane and motifs that comprise 2D quasicrystals. In this mode the algorithm explores the space of what can be made rather than the space of structures that are low in energy but not necessarily kinetically accessible.

{\em Model and learning algorithm---} We consider a class of coarse-grained model able to capture the essential physics of molecular and nanoscale self-assembly at surfaces\c{whitelam2014common}. It comprises circular particles of hard-core diameter $a$ on a two-dimensional square substrate of side $50a$. The substrate has periodic boundary conditions in both directions. Particles evolve under a stochastic dynamical protocol consisting of a grand-canonical Monte Carlo algorithm with chemical potential $\mu$, which allows particles to exchange with a notional solution\c{frenkel1996understanding}, and the virtual-move Monte Carlo algorithm\c{whitelam2009role,VMMC_3}, which allows particles to move on the surface according to an approximation of Brownian motion\c{haxton2015crystallization}. Grand-canonical moves are proposed with probability $1/(1+P)$, where $P$ is the instantaneous number of particles on the surface\c{whitelam2014common}. All trajectories start from distinct disordered configurations consisting of 500 particles randomly deposited on the surface, and are run for $t_0=10^9$ Monte Carlo steps.

The angular component of the interparticle attraction $U_{\x}(\theta)$ and the time-dependent protocol $(\mu_{\y}(t),\epsilon_{\y}(t))$ are encoded as neural networks, specified in \s{model}. Simulation potentials with angular dependence are often called ``patchy''\c{Glotzer2004patchy}. The interaction potential reflects the idea that particles interact in a complementary way, such as through DNA hybridization, hydrogen bonding, or other directional donor-acceptor mechanisms\c{swiegers2002classification,bartels2010tailoring,elemans2009molecular,pfeifer2016nano}. The control parameters $\mu$ and $\epsilon$ influence the substrate density and the strength of interparticle attractions, which could be done in experiment by e.g. varying deposition rate and temperature (we call the process of increasing $\epsilon$ ``cooling''). We denote the parameters of the neural networks by ${\bm x}$ and ${\bm y}$. Using evolutionary language we call this encoding the material's {\em genome}, an idea sketched in \f{fig_schematic}. The results of molecular simulation, using the potential and protocol defined by the genome, is the {\em phenome}.

To evolve genomes whose phenomes possess a desired property we use an iterative genetic algorithm consisting of a population dynamics combined with neuroevolution\c{GA,GA2,salimans2017evolution,montana1989training,Guber,whitelam2020learning}, specified in \s{ev}. Neuroevolution, stochastic mutation of the neural-network parameters, is equivalent in the limit of small mutations to gradient descent in the presence of Gaussian white noise\c{whitelam2020correspondence}. The learning algorithm starts in Generation 0 with a population of 100 randomly-chosen genomes, and ``expresses'' their phenomes via $t_0$ steps of the molecular simulation protocol described above. The algorithm identifies the 10 phenomes possessing the largest values of an objective function $\phi$. The 10 corresponding genomes are cloned and mutated in order to produce a new population (or generation) of 100 genomes, whose phenomes are then expressed via molecular simulation. This iterative procedure continues, generation by generation, until terminated by the user. 

The objective function or evolutionary pressure is a user-specified order parameter $\phi$, evaluated at the final time point of each simulation. In this work we consider order parameters built from two quantities. One is $C_k$, the number of clusters of interacting particles of size $k$ (called $k$-mers). The other is $N_k$, the number of convex loops of size $k$ (called $k$-gons) that can be drawn by joining the centers of interacting particles. $k$-gons are pores: designing materials with specified pore sizes is useful for e.g. gas separation\c{liu2016two}.
\begin{figure*}[] 
   \centering
\includegraphics[width=\linewidth]{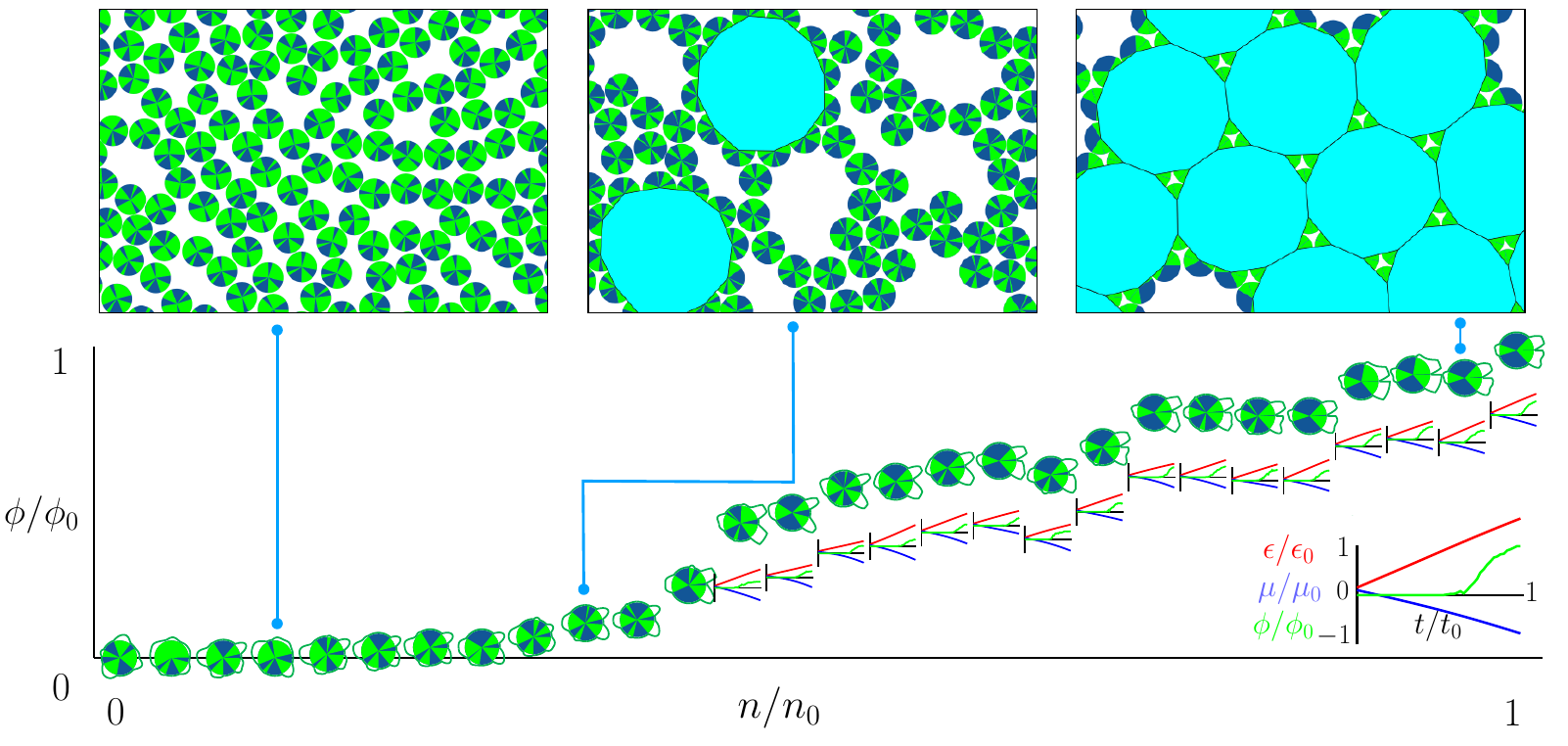} 
   \caption{\label{fig_12_gons}  Evolutionary learning directed to maximize the number of 12-gons, convex pores of 12 sides (shown light blue in images). The main panel shows the yield $\phi$, as a function of generation $n$, produced by the most successful genome (potential and protocol). The particle symbols are the data points, which also show the form of the learned potential. Below the data points are plots of the learned protocol (red and blue lines) and resulting yield (green lines) as a function of time $t$. The format of those plots is shown bottom right. Large positive values of $\epsilon$ indicate strong particle attractions, and large positive and negative values of $\mu$ promote dense and sparse substrates, respectively. The snapshots at the top show portions of a simulation box from three different generations, indicated by the blue lines. Parameters: $\phi_0=90$ 12-gons, $\epsilon_0=\mu_0=20\, \kt$, $n_0=27$ generations, $t_0=10^9$ Monte Carlo steps.}
\end{figure*}
\begin{figure}[] 
   \centering
\includegraphics[width=\linewidth]{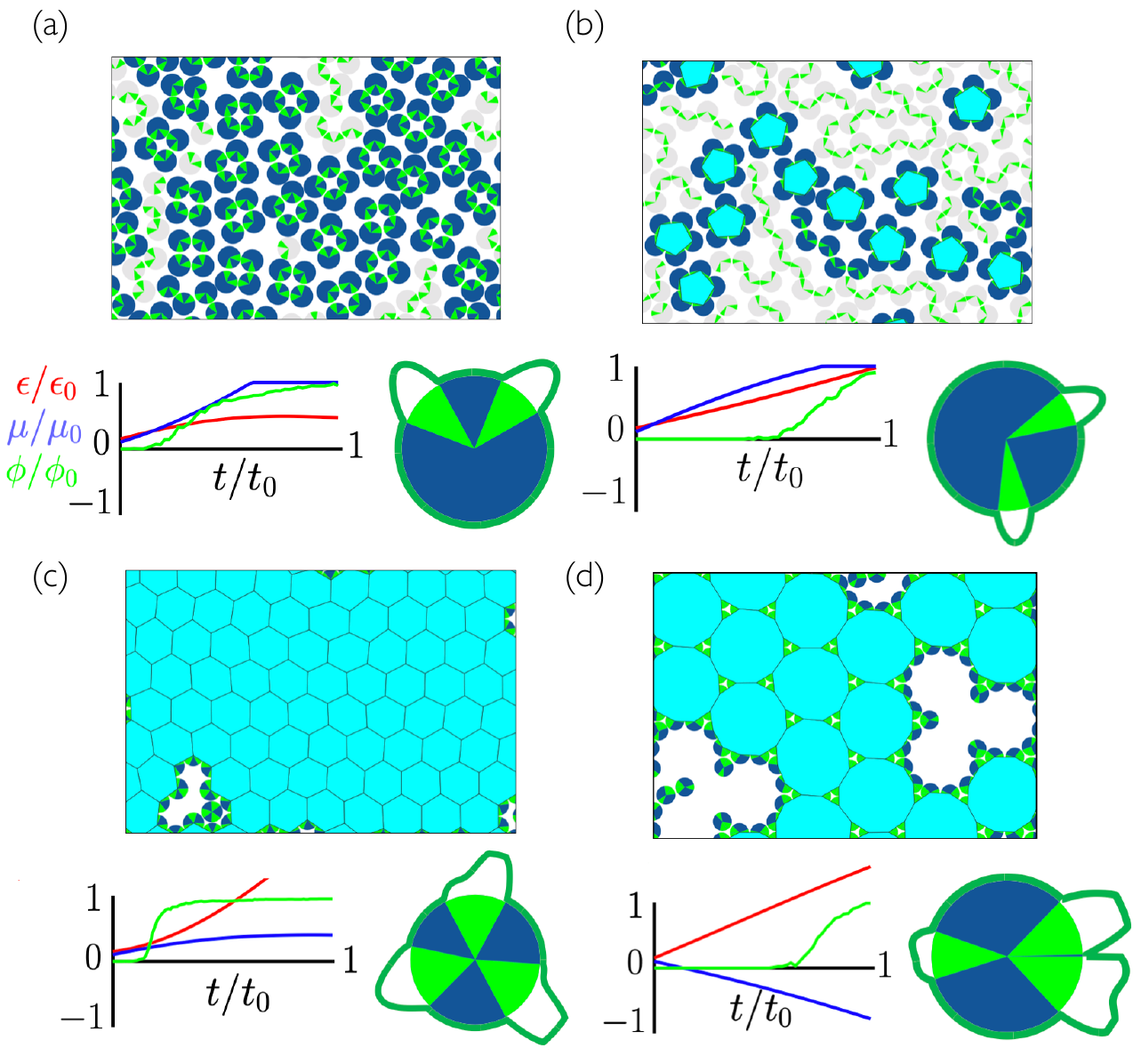} 
   \caption{\label{fig_collection} The results of evolutionary learning (genome bottom and phenome top) instructed to produce (a) 4-mers, (b) 5-mers and 5-gons, (c) 6-gons, and (d) 12-gons.}
\end{figure}

{\em Directed search---} We direct the algorithm to evolve a material containing convex pores of size 12, and set $\phi=N_{12}$. This case provides an example of an objective that is too complex to achieve without additional guidance: 12-gons are sufficiently complex that they do not form spontaneously under the random particle- and protocol design that comprises the initial stage (Generation 0) of the learning algorithm. All phenomes score zero, and the learning algorithm has nothing to work with. In this case a simple modification of the objective is sufficient to overcome the problem. We set $\phi=N_{\min(x,12)}$, where $x$ is the size of the largest convex pore seen across all 100 phenomes of a given generation. Thus if $x$ is 12 or larger then the learning algorithm selects genomes that produce 12-gons; if $x$ is smaller than 12 then it selects genomes that produce $x$-gons.

The results of several generations of learning using this objective are shown in \f{fig_12_gons}. The largest pore sizes seen in the first 4 generations are 10, 9, 11, and 11, and thereafter the first 12-gons are produced. The learning algorithm improves its design and the yield of 12-gons over evolutionary time, and eventually achieves the self-assembly of a structure dual to the 3.12.12 Archimedean tiling, which has one 3-gon and two 12-gons around each vertex\c{grunbaum1977tilings,antlanger2011stability,whitelam2016minimal}. To do so requires a sophisticated design. The particle must present sticky patches whose bisectors are separated (approximately) by angles $\pi/3$ and $5\pi/6$. In addition, the patches must be inequivalent: if all patches possess equal binding energy then kinetic traps impair the formation of the structure\c{whitelam2016minimal}. The solution identified by the learning algorithm is to make one patch weaker than the other two, and to steadily cool the substrate. The result is a hierarchical dynamics that starts with many isolated 3-gons forming from the engagement of the strong patches. Eventually the weak patches engage and cause the 3-gons to form a network, which subsequently forms 12-gons [\f{fig_12_gons_dynamics}]. The learning algorithm also evacuates the substrate, removing steric impediments to closure of the network. The resulting strategy produces a yield of 12-gons superior to that achieved by a human-designed particle and protocol [\f{fig_yield}(a)].

In \f{fig_collection} we show the results of evolutionary learning instructed to make 4-mers, 5-mers and 5-gons, and 6-gons. In each case the strategy learned is efficient: to produce 4-mers the algorithm evolves particles with two patches separated by an angle $\pi/2$, leading to compact square clusters; to make pentagonal 5-gons it evolves particles with patches separated by an angle $3 \pi/5$; and to make 6-gons the algorithm evolves particles with approximate 3-fold rotational symmetry, which self-assemble under the learned protocol into the honeycomb lattice. Similar motifs are seen in a range of real system that realize the honeycomb lattice\c{swiegers2002classification,bartels2010tailoring,elemans2009molecular,whitelam2014common}. The evolutionary pressure to achieve geometrical perfection of the interaction is relatively weak: self-assembly of a particle with perfect three-fold rotational symmetry and a learned time-dependent protocol results in a comparable yield of 6-gons [\f{fig_yield}(b)]. The evolutionary trajectories showing the emergence of these designs are shown in Figs. \ref{fig_4_mers}, \ref{fig_5_mers_gons}, and \ref{fig_6_gons}.
\begin{figure}[] 
   \centering
\includegraphics[width=\linewidth]{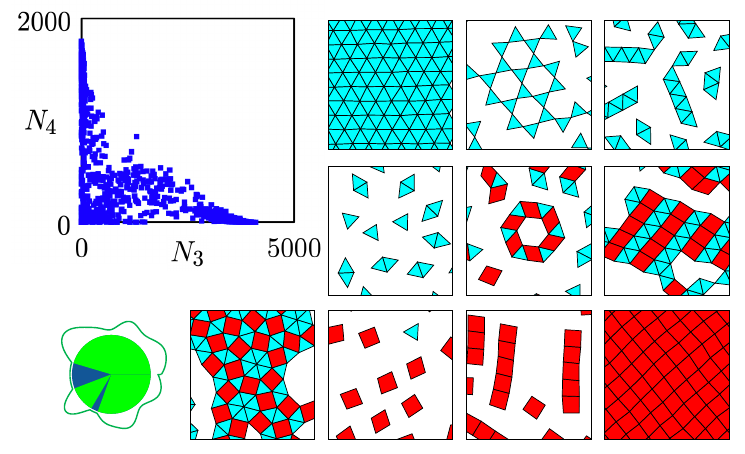} 
   \caption{\label{fig_novelty} The results of evolutionary learning instructed to produce novelty in the space of 3-gons and 4-gons. The scatterplot shows the coverage obtained in this space after 10 generations of learning; the images show examples of the structures produced (the particles underlying the polygons are not shown).}
\end{figure}

{\em Exploratory search---} We end by showing that evolutionary learning can be used in an exploratory mode, searching for novelty rather than to achieve a desired property\c{conti2018improving}. To search for novelty within the space of 4-gons and 3-gons we impose the objective function
\beq
\label{phi}
\phi=  \sum_{j} \sqrt{\frac{1}{4}\left(N_3-N_3^{(j)}\right)^2+\left(N_4-N_4^{(j)}\right)^2},
\eeq
where $j$ runs over all phenomes produced in all previous and current generations. Maximizing \eq{phi} leads to an evolutionary pressure favoring materials most unlike those produced to date, rather than materials with specified values of $N_3$ and $N_4$. Over the course of 10 generations of novelty search the learning algorithm produces the coverage of $(N_3,N_4)$ space shown in \f{fig_novelty}. Some of the polygon structures dual to the particle structures found during that exploration are shown in the figure. These include size-limited motifs; the square and triangle regular tilings of the plane; the 3.6.3.6 and 3.3.3.4.4 Archimedean tilings\c{grunbaum1977tilings,antlanger2011stability,whitelam2016minimal}; and the $\sigma$, H, and Z binding motifs prominent in dodecagonal quasicrystals\c{van2012formation}. The particle that gives rise to those motifs is shown in the figure: it has irregular six-fold symmetry, different to the five-fold and seven-fold coordination known to form similar motifs\c{van2012formation}. A larger section of the material made by this particle is shown in \f{fig_novelty2}.

{\em Conclusions---} We have shown that a neuroevolutionary learning algorithm can identify particles and protocols for the self-assembly of materials with desired properties, without input from physical principles and with no prior knowledge of self-assembly. The learning algorithm is capable of both directed and exploratory design. It can assemble a material with a user-defined property, or search for novelty in the space of specified order parameters. In the latter mode it explores the space of what can be made rather than the space of structures that are low in energy but not necessarily kinetically accessible. Moreover, the approach described here can be used to address design problems of considerable complexity: the neural-network encoding of potential and protocol extends to an arbitrary number of inputs and outputs, and evolutionary learning works with large numbers of parameters\c{Guber, salimans2017evolution}.

{\em Acknowledgments} -- This work was performed as part of a user project at the Molecular Foundry, Lawrence Berkeley National Laboratory, supported by the Office of Science, Office of Basic Energy Sciences, of the U.S. Department of Energy under Contract No. DE-AC02--05CH11231. I.T. acknowledges NSERC and performed work at the NRC under the auspices of the AI4D and MCF Programs.


\begin{thebibliography}{68}%
\makeatletter
\providecommand \@ifxundefined [1]{%
 \@ifx{#1\undefined}
}%
\providecommand \@ifnum [1]{%
 \ifnum #1\expandafter \@firstoftwo
 \else \expandafter \@secondoftwo
 \fi
}%
\providecommand \@ifx [1]{%
 \ifx #1\expandafter \@firstoftwo
 \else \expandafter \@secondoftwo
 \fi
}%
\providecommand \natexlab [1]{#1}%
\providecommand \enquote  [1]{``#1''}%
\providecommand \bibnamefont  [1]{#1}%
\providecommand \bibfnamefont [1]{#1}%
\providecommand \citenamefont [1]{#1}%
\providecommand \href@noop [0]{\@secondoftwo}%
\providecommand \href [0]{\begingroup \@sanitize@url \@href}%
\providecommand \@href[1]{\@@startlink{#1}\@@href}%
\providecommand \@@href[1]{\endgroup#1\@@endlink}%
\providecommand \@sanitize@url [0]{\catcode `\\12\catcode `\$12\catcode
  `\&12\catcode `\#12\catcode `\^12\catcode `\_12\catcode `\%12\relax}%
\providecommand \@@startlink[1]{}%
\providecommand \@@endlink[0]{}%
\providecommand \url  [0]{\begingroup\@sanitize@url \@url }%
\providecommand \@url [1]{\endgroup\@href {#1}{\urlprefix }}%
\providecommand \urlprefix  [0]{URL }%
\providecommand \Eprint [0]{\href }%
\providecommand \doibase [0]{http://dx.doi.org/}%
\providecommand \selectlanguage [0]{\@gobble}%
\providecommand \bibinfo  [0]{\@secondoftwo}%
\providecommand \bibfield  [0]{\@secondoftwo}%
\providecommand \translation [1]{[#1]}%
\providecommand \BibitemOpen [0]{}%
\providecommand \bibitemStop [0]{}%
\providecommand \bibitemNoStop [0]{.\EOS\space}%
\providecommand \EOS [0]{\spacefactor3000\relax}%
\providecommand \BibitemShut  [1]{\csname bibitem#1\endcsname}%
\let\auto@bib@innerbib\@empty
\bibitem [{\citenamefont {De~Yoreo}\ \emph {et~al.}(2016)\citenamefont
  {De~Yoreo}, \citenamefont {Mandrus}, \citenamefont {Soderholm}, \citenamefont
  {Forbes}, \citenamefont {Kanatzidis}, \citenamefont {Erlebacher},
  \citenamefont {Laskin}, \citenamefont {Wiesner}, \citenamefont {Xu},
  \citenamefont {Billinge} \emph {et~al.}}]{de2016basic}%
  \BibitemOpen
  \bibfield  {author} {\bibinfo {author} {\bibfnamefont {J.}~\bibnamefont
  {De~Yoreo}}, \bibinfo {author} {\bibfnamefont {D.}~\bibnamefont {Mandrus}},
  \bibinfo {author} {\bibfnamefont {L.}~\bibnamefont {Soderholm}}, \bibinfo
  {author} {\bibfnamefont {T.}~\bibnamefont {Forbes}}, \bibinfo {author}
  {\bibfnamefont {M.}~\bibnamefont {Kanatzidis}}, \bibinfo {author}
  {\bibfnamefont {J.}~\bibnamefont {Erlebacher}}, \bibinfo {author}
  {\bibfnamefont {J.}~\bibnamefont {Laskin}}, \bibinfo {author} {\bibfnamefont
  {U.}~\bibnamefont {Wiesner}}, \bibinfo {author} {\bibfnamefont
  {T.}~\bibnamefont {Xu}}, \bibinfo {author} {\bibfnamefont {S.}~\bibnamefont
  {Billinge}},  \emph {et~al.},\ }\href@noop {} {\emph {\bibinfo {title} {Basic
  Research Needs Workshop on Synthesis Science for Energy Relevant
  Technology}}},\ \bibinfo {type} {Tech. Rep.}\ (\bibinfo  {institution} {USDOE
  Office of Science (SC)(United States)},\ \bibinfo {year} {2016})\BibitemShut
  {NoStop}%
\bibitem [{\citenamefont {Broholm}\ \emph {et~al.}(2016)\citenamefont
  {Broholm}, \citenamefont {Fisher}, \citenamefont {Moore}, \citenamefont
  {Murnane}, \citenamefont {Moreo}, \citenamefont {Tranquada}, \citenamefont
  {Basov}, \citenamefont {Freericks}, \citenamefont {Aronson}, \citenamefont
  {MacDonald} \emph {et~al.}}]{broholm2016basic}%
  \BibitemOpen
  \bibfield  {author} {\bibinfo {author} {\bibfnamefont {C.}~\bibnamefont
  {Broholm}}, \bibinfo {author} {\bibfnamefont {I.}~\bibnamefont {Fisher}},
  \bibinfo {author} {\bibfnamefont {J.}~\bibnamefont {Moore}}, \bibinfo
  {author} {\bibfnamefont {M.}~\bibnamefont {Murnane}}, \bibinfo {author}
  {\bibfnamefont {A.}~\bibnamefont {Moreo}}, \bibinfo {author} {\bibfnamefont
  {J.}~\bibnamefont {Tranquada}}, \bibinfo {author} {\bibfnamefont
  {D.}~\bibnamefont {Basov}}, \bibinfo {author} {\bibfnamefont
  {J.}~\bibnamefont {Freericks}}, \bibinfo {author} {\bibfnamefont
  {M.}~\bibnamefont {Aronson}}, \bibinfo {author} {\bibfnamefont
  {A.}~\bibnamefont {MacDonald}},  \emph {et~al.},\ }\href@noop {} {\emph
  {\bibinfo {title} {Basic Research Needs Workshop on Quantum Materials for
  Energy Relevant Technology}}},\ \bibinfo {type} {Tech. Rep.}\ (\bibinfo
  {institution} {USDOE Office of Science (SC)(United States)},\ \bibinfo {year}
  {2016})\BibitemShut {NoStop}%
\bibitem [{\citenamefont {Cohn}\ and\ \citenamefont
  {Kumar}(2009)}]{cohn2009algorithmic}%
  \BibitemOpen
  \bibfield  {author} {\bibinfo {author} {\bibfnamefont {H.}~\bibnamefont
  {Cohn}}\ and\ \bibinfo {author} {\bibfnamefont {A.}~\bibnamefont {Kumar}},\
  }\href@noop {} {\bibfield  {journal} {\bibinfo  {journal} {Proceedings of the
  National Academy of Sciences}\ }\textbf {\bibinfo {volume} {106}},\ \bibinfo
  {pages} {9570} (\bibinfo {year} {2009})}\BibitemShut {NoStop}%
\bibitem [{\citenamefont {Torquato}(2009)}]{torquato2009inverse}%
  \BibitemOpen
  \bibfield  {author} {\bibinfo {author} {\bibfnamefont {S.}~\bibnamefont
  {Torquato}},\ }\href@noop {} {\bibfield  {journal} {\bibinfo  {journal} {Soft
  Matter}\ }\textbf {\bibinfo {volume} {5}},\ \bibinfo {pages} {1157} (\bibinfo
  {year} {2009})}\BibitemShut {NoStop}%
\bibitem [{\citenamefont {Bianchi}\ \emph {et~al.}(2012)\citenamefont
  {Bianchi}, \citenamefont {Doppelbauer}, \citenamefont {Filion}, \citenamefont
  {Dijkstra},\ and\ \citenamefont {Kahl}}]{bianchi2012predicting}%
  \BibitemOpen
  \bibfield  {author} {\bibinfo {author} {\bibfnamefont {E.}~\bibnamefont
  {Bianchi}}, \bibinfo {author} {\bibfnamefont {G.}~\bibnamefont
  {Doppelbauer}}, \bibinfo {author} {\bibfnamefont {L.}~\bibnamefont {Filion}},
  \bibinfo {author} {\bibfnamefont {M.}~\bibnamefont {Dijkstra}}, \ and\
  \bibinfo {author} {\bibfnamefont {G.}~\bibnamefont {Kahl}},\ }\href@noop {}
  {\bibfield  {journal} {\bibinfo  {journal} {The Journal of Chemical Physics}\
  }\textbf {\bibinfo {volume} {136}},\ \bibinfo {pages} {214102} (\bibinfo
  {year} {2012})}\BibitemShut {NoStop}%
\bibitem [{\citenamefont {Rechtsman}\ \emph {et~al.}(2006)\citenamefont
  {Rechtsman}, \citenamefont {Stillinger},\ and\ \citenamefont
  {Torquato}}]{rechtsman2006self}%
  \BibitemOpen
  \bibfield  {author} {\bibinfo {author} {\bibfnamefont {M.~C.}\ \bibnamefont
  {Rechtsman}}, \bibinfo {author} {\bibfnamefont {F.~H.}\ \bibnamefont
  {Stillinger}}, \ and\ \bibinfo {author} {\bibfnamefont {S.}~\bibnamefont
  {Torquato}},\ }\href@noop {} {\bibfield  {journal} {\bibinfo  {journal}
  {Physical Review E}\ }\textbf {\bibinfo {volume} {74}},\ \bibinfo {pages}
  {021404} (\bibinfo {year} {2006})}\BibitemShut {NoStop}%
\bibitem [{\citenamefont {Lindquist}\ \emph {et~al.}(2016)\citenamefont
  {Lindquist}, \citenamefont {Jadrich},\ and\ \citenamefont
  {Truskett}}]{lindquist2016communication}%
  \BibitemOpen
  \bibfield  {author} {\bibinfo {author} {\bibfnamefont {B.~A.}\ \bibnamefont
  {Lindquist}}, \bibinfo {author} {\bibfnamefont {R.~B.}\ \bibnamefont
  {Jadrich}}, \ and\ \bibinfo {author} {\bibfnamefont {T.~M.}\ \bibnamefont
  {Truskett}},\ }\href@noop {} {\bibfield  {journal} {\bibinfo  {journal}
  {Journal of Chemical Physics}\ }\textbf {\bibinfo {volume} {145}} (\bibinfo
  {year} {2016})}\BibitemShut {NoStop}%
\bibitem [{\citenamefont {Jadrich}\ \emph {et~al.}(2017)\citenamefont
  {Jadrich}, \citenamefont {Lindquist},\ and\ \citenamefont
  {Truskett}}]{jadrich2017probabilistic}%
  \BibitemOpen
  \bibfield  {author} {\bibinfo {author} {\bibfnamefont {R.}~\bibnamefont
  {Jadrich}}, \bibinfo {author} {\bibfnamefont {B.}~\bibnamefont {Lindquist}},
  \ and\ \bibinfo {author} {\bibfnamefont {T.}~\bibnamefont {Truskett}},\
  }\href@noop {} {\bibfield  {journal} {\bibinfo  {journal} {The Journal of
  Chemical Physics}\ }\textbf {\bibinfo {volume} {146}},\ \bibinfo {pages}
  {184103} (\bibinfo {year} {2017})}\BibitemShut {NoStop}%
\bibitem [{\citenamefont {Long}\ and\ \citenamefont
  {Ferguson}(2018)}]{long2018rational}%
  \BibitemOpen
  \bibfield  {author} {\bibinfo {author} {\bibfnamefont {A.~W.}\ \bibnamefont
  {Long}}\ and\ \bibinfo {author} {\bibfnamefont {A.~L.}\ \bibnamefont
  {Ferguson}},\ }\href@noop {} {\bibfield  {journal} {\bibinfo  {journal}
  {Molecular Systems Design \& Engineering}\ }\textbf {\bibinfo {volume} {3}},\
  \bibinfo {pages} {49} (\bibinfo {year} {2018})}\BibitemShut {NoStop}%
\bibitem [{\citenamefont {Gadelrab}\ \emph {et~al.}(2017)\citenamefont
  {Gadelrab}, \citenamefont {Hannon}, \citenamefont {Ross},\ and\ \citenamefont
  {Alexander-Katz}}]{gadelrab2017inverting}%
  \BibitemOpen
  \bibfield  {author} {\bibinfo {author} {\bibfnamefont {K.~R.}\ \bibnamefont
  {Gadelrab}}, \bibinfo {author} {\bibfnamefont {A.~F.}\ \bibnamefont
  {Hannon}}, \bibinfo {author} {\bibfnamefont {C.~A.}\ \bibnamefont {Ross}}, \
  and\ \bibinfo {author} {\bibfnamefont {A.}~\bibnamefont {Alexander-Katz}},\
  }\href@noop {} {\bibfield  {journal} {\bibinfo  {journal} {Molecular Systems
  Design \& Engineering}\ }\textbf {\bibinfo {volume} {2}},\ \bibinfo {pages}
  {539} (\bibinfo {year} {2017})}\BibitemShut {NoStop}%
\bibitem [{\citenamefont {Ferguson}(2017)}]{ferguson2017machine}%
  \BibitemOpen
  \bibfield  {author} {\bibinfo {author} {\bibfnamefont {A.~L.}\ \bibnamefont
  {Ferguson}},\ }\href@noop {} {\bibfield  {journal} {\bibinfo  {journal}
  {Journal of Physics: Condensed Matter}\ }\textbf {\bibinfo {volume} {30}},\
  \bibinfo {pages} {043002} (\bibinfo {year} {2017})}\BibitemShut {NoStop}%
\bibitem [{\citenamefont {Pi{\~n}eros}\ \emph {et~al.}(2018)\citenamefont
  {Pi{\~n}eros}, \citenamefont {Lindquist}, \citenamefont {Jadrich},\ and\
  \citenamefont {Truskett}}]{pineros2018inverse}%
  \BibitemOpen
  \bibfield  {author} {\bibinfo {author} {\bibfnamefont {W.~D.}\ \bibnamefont
  {Pi{\~n}eros}}, \bibinfo {author} {\bibfnamefont {B.~A.}\ \bibnamefont
  {Lindquist}}, \bibinfo {author} {\bibfnamefont {R.~B.}\ \bibnamefont
  {Jadrich}}, \ and\ \bibinfo {author} {\bibfnamefont {T.~M.}\ \bibnamefont
  {Truskett}},\ }\href@noop {} {\bibfield  {journal} {\bibinfo  {journal} {The
  Journal of Chemical Physics}\ }\textbf {\bibinfo {volume} {148}},\ \bibinfo
  {pages} {104509} (\bibinfo {year} {2018})}\BibitemShut {NoStop}%
\bibitem [{\citenamefont {van Anders}\ \emph {et~al.}(2015)\citenamefont {van
  Anders}, \citenamefont {Klotsa}, \citenamefont {Karas}, \citenamefont
  {Dodd},\ and\ \citenamefont {Glotzer}}]{van2015digital}%
  \BibitemOpen
  \bibfield  {author} {\bibinfo {author} {\bibfnamefont {G.}~\bibnamefont {van
  Anders}}, \bibinfo {author} {\bibfnamefont {D.}~\bibnamefont {Klotsa}},
  \bibinfo {author} {\bibfnamefont {A.~S.}\ \bibnamefont {Karas}}, \bibinfo
  {author} {\bibfnamefont {P.~M.}\ \bibnamefont {Dodd}}, \ and\ \bibinfo
  {author} {\bibfnamefont {S.~C.}\ \bibnamefont {Glotzer}},\ }\href@noop {}
  {\bibfield  {journal} {\bibinfo  {journal} {ACS Nano}\ }\textbf {\bibinfo
  {volume} {9}},\ \bibinfo {pages} {9542} (\bibinfo {year} {2015})}\BibitemShut
  {NoStop}%
\bibitem [{\citenamefont {Adorf}\ \emph {et~al.}(2018)\citenamefont {Adorf},
  \citenamefont {Antonaglia}, \citenamefont {Dshemuchadse},\ and\ \citenamefont
  {Glotzer}}]{adorf2018inverse}%
  \BibitemOpen
  \bibfield  {author} {\bibinfo {author} {\bibfnamefont {C.~S.}\ \bibnamefont
  {Adorf}}, \bibinfo {author} {\bibfnamefont {J.}~\bibnamefont {Antonaglia}},
  \bibinfo {author} {\bibfnamefont {J.}~\bibnamefont {Dshemuchadse}}, \ and\
  \bibinfo {author} {\bibfnamefont {S.~C.}\ \bibnamefont {Glotzer}},\
  }\href@noop {} {\bibfield  {journal} {\bibinfo  {journal} {The Journal of
  Chemical Physics}\ }\textbf {\bibinfo {volume} {149}},\ \bibinfo {pages}
  {204102} (\bibinfo {year} {2018})}\BibitemShut {NoStop}%
\bibitem [{\citenamefont {Madge}\ and\ \citenamefont
  {Miller}(2017)}]{madge2017optimising}%
  \BibitemOpen
  \bibfield  {author} {\bibinfo {author} {\bibfnamefont {J.}~\bibnamefont
  {Madge}}\ and\ \bibinfo {author} {\bibfnamefont {M.~A.}\ \bibnamefont
  {Miller}},\ }\href@noop {} {\bibfield  {journal} {\bibinfo  {journal} {Soft
  matter}\ }\textbf {\bibinfo {volume} {13}},\ \bibinfo {pages} {7780}
  (\bibinfo {year} {2017})}\BibitemShut {NoStop}%
\bibitem [{\citenamefont {Jiang}\ \emph {et~al.}(2018)\citenamefont {Jiang},
  \citenamefont {Li}, \citenamefont {Lee}, \citenamefont {Jaeger},
  \citenamefont {Heinonen},\ and\ \citenamefont
  {de~Pablo}}]{jiang2018evolutionary}%
  \BibitemOpen
  \bibfield  {author} {\bibinfo {author} {\bibfnamefont {X.}~\bibnamefont
  {Jiang}}, \bibinfo {author} {\bibfnamefont {J.}~\bibnamefont {Li}}, \bibinfo
  {author} {\bibfnamefont {V.}~\bibnamefont {Lee}}, \bibinfo {author}
  {\bibfnamefont {H.~M.}\ \bibnamefont {Jaeger}}, \bibinfo {author}
  {\bibfnamefont {O.~G.}\ \bibnamefont {Heinonen}}, \ and\ \bibinfo {author}
  {\bibfnamefont {J.~J.}\ \bibnamefont {de~Pablo}},\ }\href@noop {} {\bibfield
  {journal} {\bibinfo  {journal} {The Journal of Chemical Physics}\ }\textbf
  {\bibinfo {volume} {148}},\ \bibinfo {pages} {234302} (\bibinfo {year}
  {2018})}\BibitemShut {NoStop}%
\bibitem [{\citenamefont {Kumar}\ \emph {et~al.}(2019)\citenamefont {Kumar},
  \citenamefont {Coli}, \citenamefont {Dijkstra},\ and\ \citenamefont
  {Sastry}}]{kumar2019inverse}%
  \BibitemOpen
  \bibfield  {author} {\bibinfo {author} {\bibfnamefont {R.}~\bibnamefont
  {Kumar}}, \bibinfo {author} {\bibfnamefont {G.~M.}\ \bibnamefont {Coli}},
  \bibinfo {author} {\bibfnamefont {M.}~\bibnamefont {Dijkstra}}, \ and\
  \bibinfo {author} {\bibfnamefont {S.}~\bibnamefont {Sastry}},\ }\href@noop {}
  {\bibfield  {journal} {\bibinfo  {journal} {The Journal of Chemical Physics}\
  }\textbf {\bibinfo {volume} {151}},\ \bibinfo {pages} {084109} (\bibinfo
  {year} {2019})}\BibitemShut {NoStop}%
\bibitem [{\citenamefont {Zhou}\ \emph {et~al.}(2019)\citenamefont {Zhou},
  \citenamefont {Proctor}, \citenamefont {van Anders},\ and\ \citenamefont
  {Glotzer}}]{zhou2019alchemical}%
  \BibitemOpen
  \bibfield  {author} {\bibinfo {author} {\bibfnamefont {P.}~\bibnamefont
  {Zhou}}, \bibinfo {author} {\bibfnamefont {J.~C.}\ \bibnamefont {Proctor}},
  \bibinfo {author} {\bibfnamefont {G.}~\bibnamefont {van Anders}}, \ and\
  \bibinfo {author} {\bibfnamefont {S.~C.}\ \bibnamefont {Glotzer}},\
  }\href@noop {} {\bibfield  {journal} {\bibinfo  {journal} {Molecular
  Physics}\ }\textbf {\bibinfo {volume} {117}},\ \bibinfo {pages} {3968}
  (\bibinfo {year} {2019})}\BibitemShut {NoStop}%
\bibitem [{\citenamefont {Sherman}\ \emph {et~al.}(2020)\citenamefont
  {Sherman}, \citenamefont {Howard}, \citenamefont {Lindquist}, \citenamefont
  {Jadrich},\ and\ \citenamefont {Truskett}}]{sherman2020inverse}%
  \BibitemOpen
  \bibfield  {author} {\bibinfo {author} {\bibfnamefont {Z.~M.}\ \bibnamefont
  {Sherman}}, \bibinfo {author} {\bibfnamefont {M.~P.}\ \bibnamefont {Howard}},
  \bibinfo {author} {\bibfnamefont {B.~A.}\ \bibnamefont {Lindquist}}, \bibinfo
  {author} {\bibfnamefont {R.~B.}\ \bibnamefont {Jadrich}}, \ and\ \bibinfo
  {author} {\bibfnamefont {T.~M.}\ \bibnamefont {Truskett}},\ }\href@noop {}
  {\bibfield  {journal} {\bibinfo  {journal} {The Journal of Chemical Physics}\
  }\textbf {\bibinfo {volume} {152}},\ \bibinfo {pages} {140902} (\bibinfo
  {year} {2020})}\BibitemShut {NoStop}%
\bibitem [{\citenamefont {Reinhardt}\ and\ \citenamefont
  {Frenkel}(2014)}]{reinhardt2014numerical}%
  \BibitemOpen
  \bibfield  {author} {\bibinfo {author} {\bibfnamefont {A.}~\bibnamefont
  {Reinhardt}}\ and\ \bibinfo {author} {\bibfnamefont {D.}~\bibnamefont
  {Frenkel}},\ }\href@noop {} {\bibfield  {journal} {\bibinfo  {journal}
  {Physical Review Letters}\ }\textbf {\bibinfo {volume} {112}},\ \bibinfo
  {pages} {238103} (\bibinfo {year} {2014})}\BibitemShut {NoStop}%
\bibitem [{\citenamefont {Romano}\ \emph {et~al.}(2020)\citenamefont {Romano},
  \citenamefont {Russo}, \citenamefont {Kroc},\ and\ \citenamefont
  {{\v{S}}ulc}}]{romano2020designing}%
  \BibitemOpen
  \bibfield  {author} {\bibinfo {author} {\bibfnamefont {F.}~\bibnamefont
  {Romano}}, \bibinfo {author} {\bibfnamefont {J.}~\bibnamefont {Russo}},
  \bibinfo {author} {\bibfnamefont {L.}~\bibnamefont {Kroc}}, \ and\ \bibinfo
  {author} {\bibfnamefont {P.}~\bibnamefont {{\v{S}}ulc}},\ }\href@noop {}
  {\bibfield  {journal} {\bibinfo  {journal} {Physical Review Letters}\
  }\textbf {\bibinfo {volume} {125}},\ \bibinfo {pages} {118003} (\bibinfo
  {year} {2020})}\BibitemShut {NoStop}%
\bibitem [{\citenamefont {Miskin}\ \emph {et~al.}(2016)\citenamefont {Miskin},
  \citenamefont {Khaira}, \citenamefont {de~Pablo},\ and\ \citenamefont
  {Jaeger}}]{miskin2016turning}%
  \BibitemOpen
  \bibfield  {author} {\bibinfo {author} {\bibfnamefont {M.~Z.}\ \bibnamefont
  {Miskin}}, \bibinfo {author} {\bibfnamefont {G.}~\bibnamefont {Khaira}},
  \bibinfo {author} {\bibfnamefont {J.~J.}\ \bibnamefont {de~Pablo}}, \ and\
  \bibinfo {author} {\bibfnamefont {H.~M.}\ \bibnamefont {Jaeger}},\
  }\href@noop {} {\bibfield  {journal} {\bibinfo  {journal} {Proceedings of the
  National Academy of Sciences}\ }\textbf {\bibinfo {volume} {113}},\ \bibinfo
  {pages} {34} (\bibinfo {year} {2016})}\BibitemShut {NoStop}%
\bibitem [{\citenamefont {Klotsa}\ and\ \citenamefont
  {Jack}(2013)}]{klotsa2013controlling}%
  \BibitemOpen
  \bibfield  {author} {\bibinfo {author} {\bibfnamefont {D.}~\bibnamefont
  {Klotsa}}\ and\ \bibinfo {author} {\bibfnamefont {R.~L.}\ \bibnamefont
  {Jack}},\ }\href@noop {} {\bibfield  {journal} {\bibinfo  {journal} {The
  Journal of Chemical Physics}\ }\textbf {\bibinfo {volume} {138}},\ \bibinfo
  {pages} {094502} (\bibinfo {year} {2013})}\BibitemShut {NoStop}%
\bibitem [{\citenamefont {Raccuglia}\ \emph {et~al.}(2016)\citenamefont
  {Raccuglia}, \citenamefont {Elbert}, \citenamefont {Adler}, \citenamefont
  {Falk}, \citenamefont {Wenny}, \citenamefont {Mollo}, \citenamefont {Zeller},
  \citenamefont {Friedler}, \citenamefont {Schrier},\ and\ \citenamefont
  {Norquist}}]{raccuglia2016machine}%
  \BibitemOpen
  \bibfield  {author} {\bibinfo {author} {\bibfnamefont {P.}~\bibnamefont
  {Raccuglia}}, \bibinfo {author} {\bibfnamefont {K.~C.}\ \bibnamefont
  {Elbert}}, \bibinfo {author} {\bibfnamefont {P.~D.}\ \bibnamefont {Adler}},
  \bibinfo {author} {\bibfnamefont {C.}~\bibnamefont {Falk}}, \bibinfo {author}
  {\bibfnamefont {M.~B.}\ \bibnamefont {Wenny}}, \bibinfo {author}
  {\bibfnamefont {A.}~\bibnamefont {Mollo}}, \bibinfo {author} {\bibfnamefont
  {M.}~\bibnamefont {Zeller}}, \bibinfo {author} {\bibfnamefont {S.~A.}\
  \bibnamefont {Friedler}}, \bibinfo {author} {\bibfnamefont {J.}~\bibnamefont
  {Schrier}}, \ and\ \bibinfo {author} {\bibfnamefont {A.~J.}\ \bibnamefont
  {Norquist}},\ }\href@noop {} {\bibfield  {journal} {\bibinfo  {journal}
  {Nature}\ }\textbf {\bibinfo {volume} {533}},\ \bibinfo {pages} {73}
  (\bibinfo {year} {2016})}\BibitemShut {NoStop}%
\bibitem [{\citenamefont {Whitelam}\ and\ \citenamefont
  {Tamblyn}(2020)}]{whitelam2020learning}%
  \BibitemOpen
  \bibfield  {author} {\bibinfo {author} {\bibfnamefont {S.}~\bibnamefont
  {Whitelam}}\ and\ \bibinfo {author} {\bibfnamefont {I.}~\bibnamefont
  {Tamblyn}},\ }\href@noop {} {\bibfield  {journal} {\bibinfo  {journal}
  {Physical Review E}\ }\textbf {\bibinfo {volume} {101}},\ \bibinfo {pages}
  {052604} (\bibinfo {year} {2020})}\BibitemShut {NoStop}%
\bibitem [{\citenamefont {Tang}\ \emph {et~al.}(2016)\citenamefont {Tang},
  \citenamefont {Rupp}, \citenamefont {Yang}, \citenamefont {Edwards},
  \citenamefont {Grover},\ and\ \citenamefont {Bevan}}]{tang2016optimal}%
  \BibitemOpen
  \bibfield  {author} {\bibinfo {author} {\bibfnamefont {X.}~\bibnamefont
  {Tang}}, \bibinfo {author} {\bibfnamefont {B.}~\bibnamefont {Rupp}}, \bibinfo
  {author} {\bibfnamefont {Y.}~\bibnamefont {Yang}}, \bibinfo {author}
  {\bibfnamefont {T.~D.}\ \bibnamefont {Edwards}}, \bibinfo {author}
  {\bibfnamefont {M.~A.}\ \bibnamefont {Grover}}, \ and\ \bibinfo {author}
  {\bibfnamefont {M.~A.}\ \bibnamefont {Bevan}},\ }\href@noop {} {\bibfield
  {journal} {\bibinfo  {journal} {ACS Nano}\ }\textbf {\bibinfo {volume}
  {10}},\ \bibinfo {pages} {6791} (\bibinfo {year} {2016})}\BibitemShut
  {NoStop}%
\bibitem [{\citenamefont {Holland}(1992)}]{GA}%
  \BibitemOpen
  \bibfield  {author} {\bibinfo {author} {\bibfnamefont {J.~H.}\ \bibnamefont
  {Holland}},\ }\href@noop {} {\bibfield  {journal} {\bibinfo  {journal}
  {Scientific american}\ }\textbf {\bibinfo {volume} {267}},\ \bibinfo {pages}
  {66} (\bibinfo {year} {1992})}\BibitemShut {NoStop}%
\bibitem [{\citenamefont {Bartels}(2010)}]{bartels2010tailoring}%
  \BibitemOpen
  \bibfield  {author} {\bibinfo {author} {\bibfnamefont {L.}~\bibnamefont
  {Bartels}},\ }\href@noop {} {\bibfield  {journal} {\bibinfo  {journal}
  {Nature Chemistry}\ }\textbf {\bibinfo {volume} {2}},\ \bibinfo {pages} {87}
  (\bibinfo {year} {2010})}\BibitemShut {NoStop}%
\bibitem [{\citenamefont {Elemans}\ \emph {et~al.}(2009)\citenamefont
  {Elemans}, \citenamefont {Lei},\ and\ \citenamefont
  {De~Feyter}}]{elemans2009molecular}%
  \BibitemOpen
  \bibfield  {author} {\bibinfo {author} {\bibfnamefont {J.~A.~A.~W.}\
  \bibnamefont {Elemans}}, \bibinfo {author} {\bibfnamefont {S.}~\bibnamefont
  {Lei}}, \ and\ \bibinfo {author} {\bibfnamefont {S.}~\bibnamefont
  {De~Feyter}},\ }\href@noop {} {\bibfield  {journal} {\bibinfo  {journal}
  {Angewandte Chemie International Edition}\ }\textbf {\bibinfo {volume}
  {48}},\ \bibinfo {pages} {7298} (\bibinfo {year} {2009})}\BibitemShut
  {NoStop}%
\bibitem [{\citenamefont {Swiegers}\ and\ \citenamefont
  {Malefetse}(2002)}]{swiegers2002classification}%
  \BibitemOpen
  \bibfield  {author} {\bibinfo {author} {\bibfnamefont {G.~F.}\ \bibnamefont
  {Swiegers}}\ and\ \bibinfo {author} {\bibfnamefont {T.~J.}\ \bibnamefont
  {Malefetse}},\ }\href@noop {} {\bibfield  {journal} {\bibinfo  {journal}
  {Coordination chemistry reviews}\ }\textbf {\bibinfo {volume} {225}},\
  \bibinfo {pages} {91} (\bibinfo {year} {2002})}\BibitemShut {NoStop}%
\bibitem [{\citenamefont {Doye}\ \emph {et~al.}(2004)\citenamefont {Doye},
  \citenamefont {Louis},\ and\ \citenamefont
  {Vendruscolo}}]{doye2004inhibition}%
  \BibitemOpen
  \bibfield  {author} {\bibinfo {author} {\bibfnamefont {J.~P.~K.}\
  \bibnamefont {Doye}}, \bibinfo {author} {\bibfnamefont {A.~A.}\ \bibnamefont
  {Louis}}, \ and\ \bibinfo {author} {\bibfnamefont {M.}~\bibnamefont
  {Vendruscolo}},\ }\href@noop {} {\bibfield  {journal} {\bibinfo  {journal}
  {Physical Biology}\ }\textbf {\bibinfo {volume} {1}},\ \bibinfo {pages} {P9}
  (\bibinfo {year} {2004})}\BibitemShut {NoStop}%
\bibitem [{\citenamefont {Hagan}\ and\ \citenamefont
  {Chandler}(2006)}]{hagan2006dynamic}%
  \BibitemOpen
  \bibfield  {author} {\bibinfo {author} {\bibfnamefont {M.~F.}\ \bibnamefont
  {Hagan}}\ and\ \bibinfo {author} {\bibfnamefont {D.}~\bibnamefont
  {Chandler}},\ }\href@noop {} {\bibfield  {journal} {\bibinfo  {journal}
  {Biophysical Journal}\ }\textbf {\bibinfo {volume} {91}},\ \bibinfo {pages}
  {42} (\bibinfo {year} {2006})}\BibitemShut {NoStop}%
\bibitem [{\citenamefont {Molinero}\ and\ \citenamefont
  {Moore}(2008)}]{molinero2008water}%
  \BibitemOpen
  \bibfield  {author} {\bibinfo {author} {\bibfnamefont {V.}~\bibnamefont
  {Molinero}}\ and\ \bibinfo {author} {\bibfnamefont {E.~B.}\ \bibnamefont
  {Moore}},\ }\href@noop {} {\bibfield  {journal} {\bibinfo  {journal} {The
  Journal of Physical Chemistry B}\ }\textbf {\bibinfo {volume} {113}},\
  \bibinfo {pages} {4008} (\bibinfo {year} {2008})}\BibitemShut {NoStop}%
\bibitem [{\citenamefont {Romano}\ and\ \citenamefont
  {Sciortino}(2011)}]{romano2011colloidal}%
  \BibitemOpen
  \bibfield  {author} {\bibinfo {author} {\bibfnamefont {F.}~\bibnamefont
  {Romano}}\ and\ \bibinfo {author} {\bibfnamefont {F.}~\bibnamefont
  {Sciortino}},\ }\href@noop {} {\bibfield  {journal} {\bibinfo  {journal}
  {Nature materials}\ }\textbf {\bibinfo {volume} {10}},\ \bibinfo {pages}
  {171} (\bibinfo {year} {2011})}\BibitemShut {NoStop}%
\bibitem [{\citenamefont {Glotzer}\ \emph {et~al.}(2004)\citenamefont
  {Glotzer}, \citenamefont {Solomon},\ and\ \citenamefont
  {Kotov}}]{glotzer2004self}%
  \BibitemOpen
  \bibfield  {author} {\bibinfo {author} {\bibfnamefont {S.}~\bibnamefont
  {Glotzer}}, \bibinfo {author} {\bibfnamefont {M.}~\bibnamefont {Solomon}}, \
  and\ \bibinfo {author} {\bibfnamefont {N.~A.}\ \bibnamefont {Kotov}},\
  }\href@noop {} {\bibfield  {journal} {\bibinfo  {journal} {AIChE Journal}\
  }\textbf {\bibinfo {volume} {50}},\ \bibinfo {pages} {2978} (\bibinfo {year}
  {2004})}\BibitemShut {NoStop}%
\bibitem [{\citenamefont {Doye}\ \emph {et~al.}(2007)\citenamefont {Doye},
  \citenamefont {Louis}, \citenamefont {Lin}, \citenamefont {Allen},
  \citenamefont {Noya}, \citenamefont {Wilber}, \citenamefont {Kok},\ and\
  \citenamefont {Lyus}}]{doye2007condensed}%
  \BibitemOpen
  \bibfield  {author} {\bibinfo {author} {\bibfnamefont {J.~P.~K.}\
  \bibnamefont {Doye}}, \bibinfo {author} {\bibfnamefont {A.~A.}\ \bibnamefont
  {Louis}}, \bibinfo {author} {\bibfnamefont {I.~C.}\ \bibnamefont {Lin}},
  \bibinfo {author} {\bibfnamefont {L.~R.}\ \bibnamefont {Allen}}, \bibinfo
  {author} {\bibfnamefont {E.~G.}\ \bibnamefont {Noya}}, \bibinfo {author}
  {\bibfnamefont {A.~W.}\ \bibnamefont {Wilber}}, \bibinfo {author}
  {\bibfnamefont {H.~C.}\ \bibnamefont {Kok}}, \ and\ \bibinfo {author}
  {\bibfnamefont {R.}~\bibnamefont {Lyus}},\ }\href@noop {} {\bibfield
  {journal} {\bibinfo  {journal} {Physical Chemistry Chemical Physics}\
  }\textbf {\bibinfo {volume} {9}},\ \bibinfo {pages} {2197} (\bibinfo {year}
  {2007})}\BibitemShut {NoStop}%
\bibitem [{\citenamefont {Rapaport}(2010)}]{rapaport2010modeling}%
  \BibitemOpen
  \bibfield  {author} {\bibinfo {author} {\bibfnamefont {D.~C.}\ \bibnamefont
  {Rapaport}},\ }\href@noop {} {\bibfield  {journal} {\bibinfo  {journal}
  {Phys. Biol.}\ }\textbf {\bibinfo {volume} {7}},\ \bibinfo {pages} {045001}
  (\bibinfo {year} {2010})}\BibitemShut {NoStop}%
\bibitem [{\citenamefont {Murugan}\ \emph {et~al.}(2015)\citenamefont
  {Murugan}, \citenamefont {Zou},\ and\ \citenamefont
  {Brenner}}]{murugan2015undesired}%
  \BibitemOpen
  \bibfield  {author} {\bibinfo {author} {\bibfnamefont {A.}~\bibnamefont
  {Murugan}}, \bibinfo {author} {\bibfnamefont {J.}~\bibnamefont {Zou}}, \ and\
  \bibinfo {author} {\bibfnamefont {M.~P.}\ \bibnamefont {Brenner}},\
  }\href@noop {} {\bibfield  {journal} {\bibinfo  {journal} {Nature
  Communications}\ }\textbf {\bibinfo {volume} {6}} (\bibinfo {year}
  {2015})}\BibitemShut {NoStop}%
\bibitem [{\citenamefont {Whitelam}\ and\ \citenamefont
  {Jack}(2015)}]{whitelam2015statistical}%
  \BibitemOpen
  \bibfield  {author} {\bibinfo {author} {\bibfnamefont {S.}~\bibnamefont
  {Whitelam}}\ and\ \bibinfo {author} {\bibfnamefont {R.~L.}\ \bibnamefont
  {Jack}},\ }\href@noop {} {\bibfield  {journal} {\bibinfo  {journal} {Annual
  Review of Physical Chemistry}\ }\textbf {\bibinfo {volume} {66}},\ \bibinfo
  {pages} {143} (\bibinfo {year} {2015})}\BibitemShut {NoStop}%
\bibitem [{\citenamefont {Grunwald}\ and\ \citenamefont
  {Geissler}(2014)}]{grunwald2014patterns}%
  \BibitemOpen
  \bibfield  {author} {\bibinfo {author} {\bibfnamefont {M.}~\bibnamefont
  {Grunwald}}\ and\ \bibinfo {author} {\bibfnamefont {P.~L.}\ \bibnamefont
  {Geissler}},\ }\href@noop {} {\bibfield  {journal} {\bibinfo  {journal} {ACS
  Nano}\ }\textbf {\bibinfo {volume} {8}},\ \bibinfo {pages} {5891} (\bibinfo
  {year} {2014})}\BibitemShut {NoStop}%
\bibitem [{\citenamefont {Nguyen}\ and\ \citenamefont
  {Vaikuntanathan}(2016)}]{nguyen2016design}%
  \BibitemOpen
  \bibfield  {author} {\bibinfo {author} {\bibfnamefont {M.}~\bibnamefont
  {Nguyen}}\ and\ \bibinfo {author} {\bibfnamefont {S.}~\bibnamefont
  {Vaikuntanathan}},\ }\href@noop {} {\bibfield  {journal} {\bibinfo  {journal}
  {Proceedings of the National Academy of Sciences}\ }\textbf {\bibinfo
  {volume} {113}},\ \bibinfo {pages} {14231} (\bibinfo {year}
  {2016})}\BibitemShut {NoStop}%
\bibitem [{\citenamefont {Lutsko}(2019)}]{lutsko2019crystals}%
  \BibitemOpen
  \bibfield  {author} {\bibinfo {author} {\bibfnamefont {J.~F.}\ \bibnamefont
  {Lutsko}},\ }\href@noop {} {\bibfield  {journal} {\bibinfo  {journal}
  {Science advances}\ }\textbf {\bibinfo {volume} {5}},\ \bibinfo {pages}
  {eaav7399} (\bibinfo {year} {2019})}\BibitemShut {NoStop}%
\bibitem [{\citenamefont {Fan}\ and\ \citenamefont
  {Grunwald}(2019)}]{fan2019orientational}%
  \BibitemOpen
  \bibfield  {author} {\bibinfo {author} {\bibfnamefont {Z.}~\bibnamefont
  {Fan}}\ and\ \bibinfo {author} {\bibfnamefont {M.}~\bibnamefont {Grunwald}},\
  }\href@noop {} {\bibfield  {journal} {\bibinfo  {journal} {Journal of the
  American Chemical Society}\ }\textbf {\bibinfo {volume} {141}},\ \bibinfo
  {pages} {1980} (\bibinfo {year} {2019})}\BibitemShut {NoStop}%
\bibitem [{\citenamefont {Carpenter}\ and\ \citenamefont
  {Grunwald}(2020)}]{carpenter2020heterogeneous}%
  \BibitemOpen
  \bibfield  {author} {\bibinfo {author} {\bibfnamefont {J.~E.}\ \bibnamefont
  {Carpenter}}\ and\ \bibinfo {author} {\bibfnamefont {M.}~\bibnamefont
  {Grunwald}},\ }\href@noop {} {\bibfield  {journal} {\bibinfo  {journal}
  {Journal of the American Chemical Society}\ }\textbf {\bibinfo {volume}
  {142}},\ \bibinfo {pages} {10755} (\bibinfo {year} {2020})}\BibitemShut
  {NoStop}%
\bibitem [{\citenamefont {Thorkelsson}\ \emph {et~al.}(2015)\citenamefont
  {Thorkelsson}, \citenamefont {Bai},\ and\ \citenamefont
  {Xu}}]{thorkelsson2015self}%
  \BibitemOpen
  \bibfield  {author} {\bibinfo {author} {\bibfnamefont {K.}~\bibnamefont
  {Thorkelsson}}, \bibinfo {author} {\bibfnamefont {P.}~\bibnamefont {Bai}}, \
  and\ \bibinfo {author} {\bibfnamefont {T.}~\bibnamefont {Xu}},\ }\href@noop
  {} {\bibfield  {journal} {\bibinfo  {journal} {Nano Today}\ }\textbf
  {\bibinfo {volume} {10}},\ \bibinfo {pages} {48} (\bibinfo {year}
  {2015})}\BibitemShut {NoStop}%
\bibitem [{\citenamefont {Biancaniello}\ \emph {et~al.}(2005)\citenamefont
  {Biancaniello}, \citenamefont {Kim},\ and\ \citenamefont
  {Crocker}}]{biancaniello2005colloidal}%
  \BibitemOpen
  \bibfield  {author} {\bibinfo {author} {\bibfnamefont {P.~L.}\ \bibnamefont
  {Biancaniello}}, \bibinfo {author} {\bibfnamefont {A.~J.}\ \bibnamefont
  {Kim}}, \ and\ \bibinfo {author} {\bibfnamefont {J.~C.}\ \bibnamefont
  {Crocker}},\ }\href@noop {} {\bibfield  {journal} {\bibinfo  {journal}
  {Physical Review Letters}\ }\textbf {\bibinfo {volume} {94}},\ \bibinfo
  {pages} {058302} (\bibinfo {year} {2005})}\BibitemShut {NoStop}%
\bibitem [{\citenamefont {Park}\ \emph {et~al.}(2008)\citenamefont {Park},
  \citenamefont {Lytton-Jean}, \citenamefont {Lee}, \citenamefont {Weigand},
  \citenamefont {Schatz},\ and\ \citenamefont {Mirkin}}]{park2008dna}%
  \BibitemOpen
  \bibfield  {author} {\bibinfo {author} {\bibfnamefont {S.~Y.}\ \bibnamefont
  {Park}}, \bibinfo {author} {\bibfnamefont {A.~K.}\ \bibnamefont
  {Lytton-Jean}}, \bibinfo {author} {\bibfnamefont {B.}~\bibnamefont {Lee}},
  \bibinfo {author} {\bibfnamefont {S.}~\bibnamefont {Weigand}}, \bibinfo
  {author} {\bibfnamefont {G.~C.}\ \bibnamefont {Schatz}}, \ and\ \bibinfo
  {author} {\bibfnamefont {C.~A.}\ \bibnamefont {Mirkin}},\ }\href@noop {}
  {\bibfield  {journal} {\bibinfo  {journal} {Nature}\ }\textbf {\bibinfo
  {volume} {451}},\ \bibinfo {pages} {553} (\bibinfo {year}
  {2008})}\BibitemShut {NoStop}%
\bibitem [{\citenamefont {Nykypanchuk}\ \emph {et~al.}(2008)\citenamefont
  {Nykypanchuk}, \citenamefont {Maye}, \citenamefont {van~der Lelie},\ and\
  \citenamefont {Gang}}]{nykypanchuk2008dna}%
  \BibitemOpen
  \bibfield  {author} {\bibinfo {author} {\bibfnamefont {D.}~\bibnamefont
  {Nykypanchuk}}, \bibinfo {author} {\bibfnamefont {M.~M.}\ \bibnamefont
  {Maye}}, \bibinfo {author} {\bibfnamefont {D.}~\bibnamefont {van~der Lelie}},
  \ and\ \bibinfo {author} {\bibfnamefont {O.}~\bibnamefont {Gang}},\
  }\href@noop {} {\bibfield  {journal} {\bibinfo  {journal} {Nature}\ }\textbf
  {\bibinfo {volume} {451}},\ \bibinfo {pages} {549} (\bibinfo {year}
  {2008})}\BibitemShut {NoStop}%
\bibitem [{\citenamefont {Pfeifer}\ and\ \citenamefont
  {Sacc{\`a}}(2018)}]{pfeifer2018synthetic}%
  \BibitemOpen
  \bibfield  {author} {\bibinfo {author} {\bibfnamefont {W.}~\bibnamefont
  {Pfeifer}}\ and\ \bibinfo {author} {\bibfnamefont {B.}~\bibnamefont
  {Sacc{\`a}}},\ }\href@noop {} {\bibfield  {journal} {\bibinfo  {journal}
  {Biological Chemistry}\ }\textbf {\bibinfo {volume} {399}},\ \bibinfo {pages}
  {773} (\bibinfo {year} {2018})}\BibitemShut {NoStop}%
\bibitem [{\citenamefont {De~Yoreo}\ \emph {et~al.}(2015)\citenamefont
  {De~Yoreo}, \citenamefont {Gilbert}, \citenamefont {Sommerdijk},
  \citenamefont {Penn}, \citenamefont {Whitelam}, \citenamefont {Joester},
  \citenamefont {Zhang}, \citenamefont {Rimer}, \citenamefont {Navrotsky},
  \citenamefont {Banfield} \emph {et~al.}}]{de2015crystallization}%
  \BibitemOpen
  \bibfield  {author} {\bibinfo {author} {\bibfnamefont {J.~J.}\ \bibnamefont
  {De~Yoreo}}, \bibinfo {author} {\bibfnamefont {P.~U.}\ \bibnamefont
  {Gilbert}}, \bibinfo {author} {\bibfnamefont {N.~A.}\ \bibnamefont
  {Sommerdijk}}, \bibinfo {author} {\bibfnamefont {R.~L.}\ \bibnamefont
  {Penn}}, \bibinfo {author} {\bibfnamefont {S.}~\bibnamefont {Whitelam}},
  \bibinfo {author} {\bibfnamefont {D.}~\bibnamefont {Joester}}, \bibinfo
  {author} {\bibfnamefont {H.}~\bibnamefont {Zhang}}, \bibinfo {author}
  {\bibfnamefont {J.~D.}\ \bibnamefont {Rimer}}, \bibinfo {author}
  {\bibfnamefont {A.}~\bibnamefont {Navrotsky}}, \bibinfo {author}
  {\bibfnamefont {J.~F.}\ \bibnamefont {Banfield}},  \emph {et~al.},\
  }\href@noop {} {\bibfield  {journal} {\bibinfo  {journal} {Science}\ }\textbf
  {\bibinfo {volume} {349}},\ \bibinfo {pages} {aaa6760} (\bibinfo {year}
  {2015})}\BibitemShut {NoStop}%
\bibitem [{\citenamefont {Whitelam}\ \emph {et~al.}(2014)\citenamefont
  {Whitelam}, \citenamefont {Tamblyn}, \citenamefont {Haxton}, \citenamefont
  {Wieland}, \citenamefont {Champness}, \citenamefont {Garrahan},\ and\
  \citenamefont {Beton}}]{whitelam2014common}%
  \BibitemOpen
  \bibfield  {author} {\bibinfo {author} {\bibfnamefont {S.}~\bibnamefont
  {Whitelam}}, \bibinfo {author} {\bibfnamefont {I.}~\bibnamefont {Tamblyn}},
  \bibinfo {author} {\bibfnamefont {T.~K.}\ \bibnamefont {Haxton}}, \bibinfo
  {author} {\bibfnamefont {M.~B.}\ \bibnamefont {Wieland}}, \bibinfo {author}
  {\bibfnamefont {N.~R.}\ \bibnamefont {Champness}}, \bibinfo {author}
  {\bibfnamefont {J.~P.}\ \bibnamefont {Garrahan}}, \ and\ \bibinfo {author}
  {\bibfnamefont {P.~H.}\ \bibnamefont {Beton}},\ }\href@noop {} {\bibfield
  {journal} {\bibinfo  {journal} {Physical Review X}\ }\textbf {\bibinfo
  {volume} {4}},\ \bibinfo {pages} {011044} (\bibinfo {year}
  {2014})}\BibitemShut {NoStop}%
\bibitem [{\citenamefont {Fogel}\ and\ \citenamefont {Stayton}(1994)}]{GA2}%
  \BibitemOpen
  \bibfield  {author} {\bibinfo {author} {\bibfnamefont {D.~B.}\ \bibnamefont
  {Fogel}}\ and\ \bibinfo {author} {\bibfnamefont {L.~C.}\ \bibnamefont
  {Stayton}},\ }\href@noop {} {\bibfield  {journal} {\bibinfo  {journal}
  {BioSystems}\ }\textbf {\bibinfo {volume} {32}},\ \bibinfo {pages} {171}
  (\bibinfo {year} {1994})}\BibitemShut {NoStop}%
\bibitem [{\citenamefont {Salimans}\ \emph {et~al.}(2017)\citenamefont
  {Salimans}, \citenamefont {Ho}, \citenamefont {Chen}, \citenamefont {Sidor},\
  and\ \citenamefont {Sutskever}}]{salimans2017evolution}%
  \BibitemOpen
  \bibfield  {author} {\bibinfo {author} {\bibfnamefont {T.}~\bibnamefont
  {Salimans}}, \bibinfo {author} {\bibfnamefont {J.}~\bibnamefont {Ho}},
  \bibinfo {author} {\bibfnamefont {X.}~\bibnamefont {Chen}}, \bibinfo {author}
  {\bibfnamefont {S.}~\bibnamefont {Sidor}}, \ and\ \bibinfo {author}
  {\bibfnamefont {I.}~\bibnamefont {Sutskever}},\ }\href@noop {} {\bibfield
  {journal} {\bibinfo  {journal} {arXiv preprint arXiv:1703.03864}\ } (\bibinfo
  {year} {2017})}\BibitemShut {NoStop}%
\bibitem [{\citenamefont {Montana}\ and\ \citenamefont
  {Davis}(1989)}]{montana1989training}%
  \BibitemOpen
  \bibfield  {author} {\bibinfo {author} {\bibfnamefont {D.~J.}\ \bibnamefont
  {Montana}}\ and\ \bibinfo {author} {\bibfnamefont {L.}~\bibnamefont
  {Davis}},\ }in\ \href@noop {} {\emph {\bibinfo {booktitle} {IJCAI}}},\
  Vol.~\bibinfo {volume} {89}\ (\bibinfo {year} {1989})\ pp.\ \bibinfo {pages}
  {762--767}\BibitemShut {NoStop}%
\bibitem [{\citenamefont {Such}\ \emph {et~al.}(2017)\citenamefont {Such},
  \citenamefont {Madhavan}, \citenamefont {Conti}, \citenamefont {Lehman},
  \citenamefont {Stanley},\ and\ \citenamefont {Clune}}]{Guber}%
  \BibitemOpen
  \bibfield  {author} {\bibinfo {author} {\bibfnamefont {F.~P.}\ \bibnamefont
  {Such}}, \bibinfo {author} {\bibfnamefont {V.}~\bibnamefont {Madhavan}},
  \bibinfo {author} {\bibfnamefont {E.}~\bibnamefont {Conti}}, \bibinfo
  {author} {\bibfnamefont {J.}~\bibnamefont {Lehman}}, \bibinfo {author}
  {\bibfnamefont {K.~O.}\ \bibnamefont {Stanley}}, \ and\ \bibinfo {author}
  {\bibfnamefont {J.}~\bibnamefont {Clune}},\ }\href@noop {} {\bibfield
  {journal} {\bibinfo  {journal} {arXiv preprint arXiv:1712.06567}\ } (\bibinfo
  {year} {2017})}\BibitemShut {NoStop}%
\bibitem [{\citenamefont {Whitelam}\ \emph {et~al.}(2020)\citenamefont
  {Whitelam}, \citenamefont {Selin}, \citenamefont {Park},\ and\ \citenamefont
  {Tamblyn}}]{whitelam2020correspondence}%
  \BibitemOpen
  \bibfield  {author} {\bibinfo {author} {\bibfnamefont {S.}~\bibnamefont
  {Whitelam}}, \bibinfo {author} {\bibfnamefont {V.}~\bibnamefont {Selin}},
  \bibinfo {author} {\bibfnamefont {S.-W.}\ \bibnamefont {Park}}, \ and\
  \bibinfo {author} {\bibfnamefont {I.}~\bibnamefont {Tamblyn}},\ }\href@noop
  {} {\bibfield  {journal} {\bibinfo  {journal} {arXiv preprint
  arXiv:2008.06643}\ } (\bibinfo {year} {2020})}\BibitemShut {NoStop}%
\bibitem [{\citenamefont {Conti}\ \emph {et~al.}(2018)\citenamefont {Conti},
  \citenamefont {Madhavan}, \citenamefont {Such}, \citenamefont {Lehman},
  \citenamefont {Stanley},\ and\ \citenamefont {Clune}}]{conti2018improving}%
  \BibitemOpen
  \bibfield  {author} {\bibinfo {author} {\bibfnamefont {E.}~\bibnamefont
  {Conti}}, \bibinfo {author} {\bibfnamefont {V.}~\bibnamefont {Madhavan}},
  \bibinfo {author} {\bibfnamefont {F.~P.}\ \bibnamefont {Such}}, \bibinfo
  {author} {\bibfnamefont {J.}~\bibnamefont {Lehman}}, \bibinfo {author}
  {\bibfnamefont {K.}~\bibnamefont {Stanley}}, \ and\ \bibinfo {author}
  {\bibfnamefont {J.}~\bibnamefont {Clune}},\ }in\ \href@noop {} {\emph
  {\bibinfo {booktitle} {Advances in neural information processing systems}}}\
  (\bibinfo {year} {2018})\ pp.\ \bibinfo {pages} {5027--5038}\BibitemShut
  {NoStop}%
\bibitem [{\citenamefont {Frenkel}\ and\ \citenamefont
  {Smit}(1996)}]{frenkel1996understanding}%
  \BibitemOpen
  \bibfield  {author} {\bibinfo {author} {\bibfnamefont {D.}~\bibnamefont
  {Frenkel}}\ and\ \bibinfo {author} {\bibfnamefont {B.}~\bibnamefont {Smit}},\
  }\href@noop {} {\emph {\bibinfo {title} {{Understanding Molecular Simulation:
  From Algorithms to Applications}}}}\ (\bibinfo  {publisher} {Academic Press,
  Inc. Orlando, FL, USA},\ \bibinfo {year} {1996})\BibitemShut {NoStop}%
\bibitem [{\citenamefont {Whitelam}\ \emph {et~al.}(2009)\citenamefont
  {Whitelam}, \citenamefont {Feng}, \citenamefont {Hagan},\ and\ \citenamefont
  {Geissler}}]{whitelam2009role}%
  \BibitemOpen
  \bibfield  {author} {\bibinfo {author} {\bibfnamefont {S.}~\bibnamefont
  {Whitelam}}, \bibinfo {author} {\bibfnamefont {E.~H.}\ \bibnamefont {Feng}},
  \bibinfo {author} {\bibfnamefont {M.~F.}\ \bibnamefont {Hagan}}, \ and\
  \bibinfo {author} {\bibfnamefont {P.~L.}\ \bibnamefont {Geissler}},\
  }\href@noop {} {\bibfield  {journal} {\bibinfo  {journal} {Soft Matter}\
  }\textbf {\bibinfo {volume} {5}},\ \bibinfo {pages} {1251} (\bibinfo {year}
  {2009})}\BibitemShut {NoStop}%
\bibitem [{\citenamefont {Hedges}()}]{VMMC_3}%
  \BibitemOpen
  \bibfield  {author} {\bibinfo {author} {\bibfnamefont {L.~O.}\ \bibnamefont
  {Hedges}},\ }\href {http://vmmc.xyz} {\enquote {\bibinfo {title}
  {http://vmmc.xyz},}\ }\BibitemShut {NoStop}%
\bibitem [{\citenamefont {Haxton}\ \emph {et~al.}(2015)\citenamefont {Haxton},
  \citenamefont {Hedges},\ and\ \citenamefont
  {Whitelam}}]{haxton2015crystallization}%
  \BibitemOpen
  \bibfield  {author} {\bibinfo {author} {\bibfnamefont {T.~K.}\ \bibnamefont
  {Haxton}}, \bibinfo {author} {\bibfnamefont {L.~O.}\ \bibnamefont {Hedges}},
  \ and\ \bibinfo {author} {\bibfnamefont {S.}~\bibnamefont {Whitelam}},\
  }\href@noop {} {\bibfield  {journal} {\bibinfo  {journal} {Soft matter}\
  }\textbf {\bibinfo {volume} {11}},\ \bibinfo {pages} {9307} (\bibinfo {year}
  {2015})}\BibitemShut {NoStop}%
\bibitem [{\citenamefont {Zhang}\ and\ \citenamefont
  {Glotzer}(2004)}]{Glotzer2004patchy}%
  \BibitemOpen
  \bibfield  {author} {\bibinfo {author} {\bibfnamefont {Z.}~\bibnamefont
  {Zhang}}\ and\ \bibinfo {author} {\bibfnamefont {S.~C.}\ \bibnamefont
  {Glotzer}},\ }\href@noop {} {\bibfield  {journal} {\bibinfo  {journal} {Nano
  Letters}\ }\textbf {\bibinfo {volume} {4}},\ \bibinfo {pages} {1407}
  (\bibinfo {year} {2004})}\BibitemShut {NoStop}%
\bibitem [{\citenamefont {Pfeifer}\ and\ \citenamefont
  {Sacc{\`a}}(2016)}]{pfeifer2016nano}%
  \BibitemOpen
  \bibfield  {author} {\bibinfo {author} {\bibfnamefont {W.}~\bibnamefont
  {Pfeifer}}\ and\ \bibinfo {author} {\bibfnamefont {B.}~\bibnamefont
  {Sacc{\`a}}},\ }\href@noop {} {\bibfield  {journal} {\bibinfo  {journal}
  {ChemBioChem}\ }\textbf {\bibinfo {volume} {17}},\ \bibinfo {pages} {1063}
  (\bibinfo {year} {2016})}\BibitemShut {NoStop}%
\bibitem [{\citenamefont {Liu}\ \emph {et~al.}(2016)\citenamefont {Liu},
  \citenamefont {Jin},\ and\ \citenamefont {Xu}}]{liu2016two}%
  \BibitemOpen
  \bibfield  {author} {\bibinfo {author} {\bibfnamefont {G.}~\bibnamefont
  {Liu}}, \bibinfo {author} {\bibfnamefont {W.}~\bibnamefont {Jin}}, \ and\
  \bibinfo {author} {\bibfnamefont {N.}~\bibnamefont {Xu}},\ }\href@noop {}
  {\bibfield  {journal} {\bibinfo  {journal} {Angewandte Chemie International
  Edition}\ }\textbf {\bibinfo {volume} {55}},\ \bibinfo {pages} {13384}
  (\bibinfo {year} {2016})}\BibitemShut {NoStop}%
\bibitem [{\citenamefont {Grunbaum}\ and\ \citenamefont
  {Shephard}(1977)}]{grunbaum1977tilings}%
  \BibitemOpen
  \bibfield  {author} {\bibinfo {author} {\bibfnamefont {B.}~\bibnamefont
  {Grunbaum}}\ and\ \bibinfo {author} {\bibfnamefont {G.~C.}\ \bibnamefont
  {Shephard}},\ }\href@noop {} {\bibfield  {journal} {\bibinfo  {journal}
  {Mathematics Magazine}\ }\textbf {\bibinfo {volume} {50}},\ \bibinfo {pages}
  {227} (\bibinfo {year} {1977})}\BibitemShut {NoStop}%
\bibitem [{\citenamefont {Antlanger}\ \emph {et~al.}(2011)\citenamefont
  {Antlanger}, \citenamefont {Doppelbauer},\ and\ \citenamefont
  {Kahl}}]{antlanger2011stability}%
  \BibitemOpen
  \bibfield  {author} {\bibinfo {author} {\bibfnamefont {M.}~\bibnamefont
  {Antlanger}}, \bibinfo {author} {\bibfnamefont {G.}~\bibnamefont
  {Doppelbauer}}, \ and\ \bibinfo {author} {\bibfnamefont {G.}~\bibnamefont
  {Kahl}},\ }\href@noop {} {\bibfield  {journal} {\bibinfo  {journal} {Journal
  of Physics: Condensed Matter}\ }\textbf {\bibinfo {volume} {23}},\ \bibinfo
  {pages} {404206} (\bibinfo {year} {2011})}\BibitemShut {NoStop}%
\bibitem [{\citenamefont {Whitelam}(2016)}]{whitelam2016minimal}%
  \BibitemOpen
  \bibfield  {author} {\bibinfo {author} {\bibfnamefont {S.}~\bibnamefont
  {Whitelam}},\ }\href@noop {} {\bibfield  {journal} {\bibinfo  {journal}
  {Physical Review Letters}\ }\textbf {\bibinfo {volume} {117}},\ \bibinfo
  {pages} {228003} (\bibinfo {year} {2016})}\BibitemShut {NoStop}%
\bibitem [{\citenamefont {van~der Linden}\ \emph {et~al.}(2012)\citenamefont
  {van~der Linden}, \citenamefont {Doye},\ and\ \citenamefont
  {Louis}}]{van2012formation}%
  \BibitemOpen
  \bibfield  {author} {\bibinfo {author} {\bibfnamefont {M.~N.}\ \bibnamefont
  {van~der Linden}}, \bibinfo {author} {\bibfnamefont {J.~P.}\ \bibnamefont
  {Doye}}, \ and\ \bibinfo {author} {\bibfnamefont {A.~A.}\ \bibnamefont
  {Louis}},\ }\href@noop {} {\bibfield  {journal} {\bibinfo  {journal} {The
  Journal of Chemical Physics}\ }\textbf {\bibinfo {volume} {136}},\ \bibinfo
  {pages} {054904} (\bibinfo {year} {2012})}\BibitemShut {NoStop}%
\end{thebibliography}

%

\clearpage

\renewcommand{\theequation}{S\arabic{equation}}
\renewcommand{\thefigure}{S\arabic{figure}}
\renewcommand{\thesection}{S\arabic{section}}

\setcounter{equation}{0}
\setcounter{section}{0}
\setcounter{figure}{0}

\section{Neural-network encoding of potential and protocol} 
\label{model}

Interparticle attraction and time-dependent assembly protocols are encoded as single-layer neural networks of $K=10^3$ hidden nodes, sketched in~\f{fig_schematic}. Particles possess an attractive interaction of range $a/10$. The angular component of the interaction is the radial-basis-function neural network
\beq
\label{net1}
 U_{\x}(\theta) = \sum_{\alpha=0}^{K-1} x_{3 \alpha+1} \exp\left[ -\frac{1}{2}\left(\frac{\theta-x_{3 \alpha +3}}{{\rm e}^{x_{3\alpha+2}}}\right)^2 \right],
\eeq
containing $N=3K$ trainable parameters $\x = \{x_1,\dots,x_N\}$. If $U_{\x}(\theta)$ lies outside the interval $[0,1]$ then it is set to the appropriate edge of the interval. In images, angular portions of the particle corresponding to $U_{\x}(\theta)>0$ are colored green, and are otherwise blue. Initially (in Generation 0) we set $\exp(x_{3\alpha+2})=0.2$, $x_{3 \alpha +3}=2 \pi \alpha/K$, and choose the $x_{3 \alpha+1} \sim {\cal N}(0,0.02)$ to be Gaussian random numbers. The potential remains fixed for all time within each simulation. For computational efficiency we evaluate the neural network only once, at the start of the simulation, in order to create a lookup table for $U_{\x}(\theta)$ with $\theta$ discretized over 1000 points on the interval $[0,2 \pi)$.

Two particles $i$ and $j$ whose centers are a distance $d$ apart, where $a < d \leq 11a/10$, experience an energy of interaction 
\beq
E_{ij}= -\epsilon \min(U_{\x}(\theta_{ij}),U_{\x}(\theta_{ji})).
\eeq
Here $\epsilon>0$ sets the scale of the interaction. $\theta_{ij} \in [0, 2 \pi)$ is the angle (in an anti-clockwise sense) between two lines, the line joining the center of particle $i$ to the point specified by $\theta =0$ on its circumference, and the line joining the center of particle $i$ to the center of particle $j$: see \f{fig_potential}. The minimum function encodes the idea that particles interact in a complementary way, such as through DNA hybridization, hydrogen bonding, or other directional donor-acceptor mechanisms.

The time-dependent protocol $(\mu_{\y}(t),\epsilon_{\y}(t))$ is encoded by a second neural network. Each trajectory starts with control-parameter values $\epsilon=3\, \kt$ and $\mu=2\, \kt$. 1000 times within each trajectory, at time increments of $10^{-3} t_0$, the control parameters are set to the new values $\epsilon \to \epsilon+\Delta \epsilon_{\y}(t)$ and $\mu \to \mu+ \Delta \mu_{\y}(t)$, where the neural network
\bea
\label{net2}
(\Delta \epsilon_{\y}(t),\Delta \mu_{\y}(t))&=&\frac{\kt}{K} \sum_{\alpha=0}^{K-1} (y_{4 \alpha +1},y_{4 \alpha +2}) \nonumber \\&\times& \tanh(y_{4 \alpha +3} t+y_{4 \alpha +4})
\eea
contains $M=4K$ trainable parameters $\y = \{y_1,\dots,y_N\}$. If $\mu$ moves outwith the interval $[-20,20]$ it is returned to the appropriate edge of the interval. Initially (in Generation 0) all parameters of this network are Gaussian random numbers, $y_i \sim {\cal N}(0,1)$.

\section{Evolutionary learning algorithm}
\label{ev}
 The evolutionary algorithm starts with 100 molecular simulations using distinct genomes $(\x,\y)$, randomized as described in \s{model}. This set of 100 simulations is called Generation 0, and, following molecular simulation, results in 100 phenomes. Let $\phi$ quantify the design goal (e.g. the number of clusters or pores of a certain size). The algorithm selects the 10 genomes responsible for the phenomes having the 10 largest values of $\phi$, choosing randomly in the event of equal scores. (For the novelty search used to produce \f{fig_novelty} we selected the top 25 phenomes, rather than the top 10). To create the 100 genomes that comprise Generation 1 it draws 100 times randomly with replacement from this set of 10 genomes, and mutates each by a set of Gaussian random numbers,
\beq
\x \to \x + \delta \x \quad {\rm and} \quad \y \to \y + \delta \y.
\eeq
Here $\delta \x = \{\delta x_1,\dots, \delta x_N\}$ with $\delta x_i \sim {\cal N}(0,\sigma_x^2)$, and  $\delta \y = \{\delta y_1,\dots, \delta y_M\}$ with $\delta y_i \sim {\cal N}(0,\sigma_y^2)$. The parameters $\sigma_x$ and $\sigma_y$ are chosen independently for each simulation, as the absolute value of the Gaussian random numbers ${\cal N}(0,0.1)$  and ${\cal N}(0,0.02)$, respectively. Doing so results in a combination of many small mutations and the occasional large mutation. Molecular simulation of this new set of 100 genomes results in the 100 phenomes of Generation 1, and so on.
\begin{figure}[] 
   \centering
\includegraphics[width=0.5\linewidth]{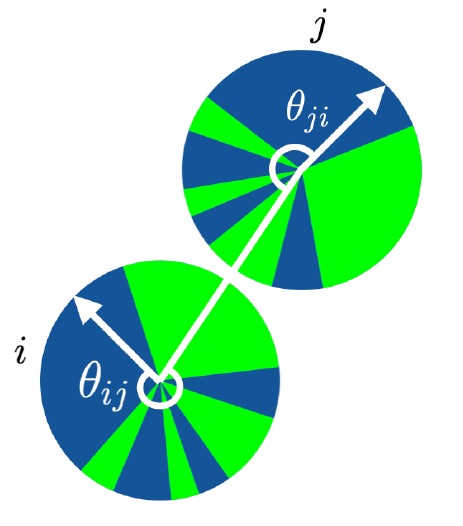} 
   \caption{\label{fig_potential} Geometry for the interaction potential. The white arrow indicates the line from the center of the particle to the point on its circumference specified by $\theta=0$.}
\end{figure}

\clearpage
\onecolumngrid
\section{Supplementary figures}
\label{figs}

\begin{figure*}[b] 
   \centering
\includegraphics[width=\linewidth]{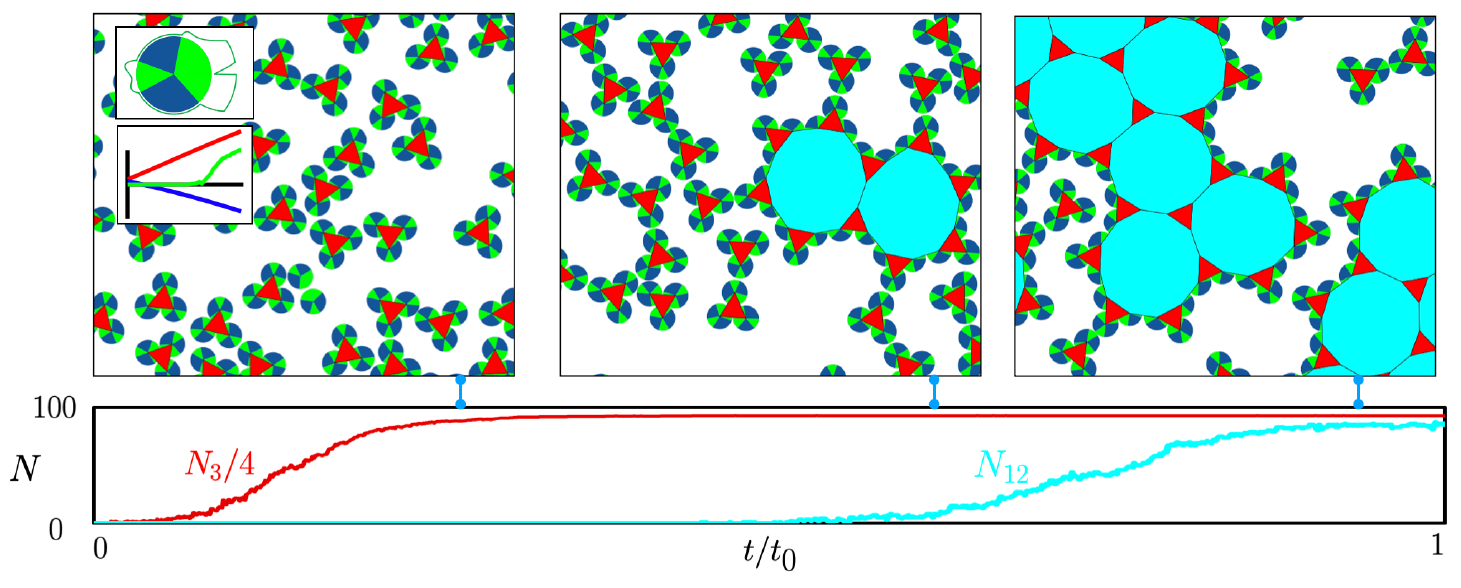} 
   \caption{\label{fig_12_gons_dynamics} Self-assembly dynamics produced by the most successful genome (potential and protocol, shown top left) from generation 27 of \f{fig_12_gons}. $N_3$ and $N_{12}$ are the number of 3-gons (red) and 12-gons (light blue). The snapshots are taken from the times indicated by the blue lines. 12-gons form relatively late in the trajectory, emphasizing that self-assembly is a ``sparse reward'' problem: the objective $\phi$ can be zero for most of the simulation. Evolutionary methods that require information from only the final time point of the simulation are natural ways to tackle such problems. Parameters: $t_0=10^9$ Monte Carlo steps.}
\end{figure*}

\clearpage
\begin{figure*}[] 
   \centering
\includegraphics[width=\linewidth]{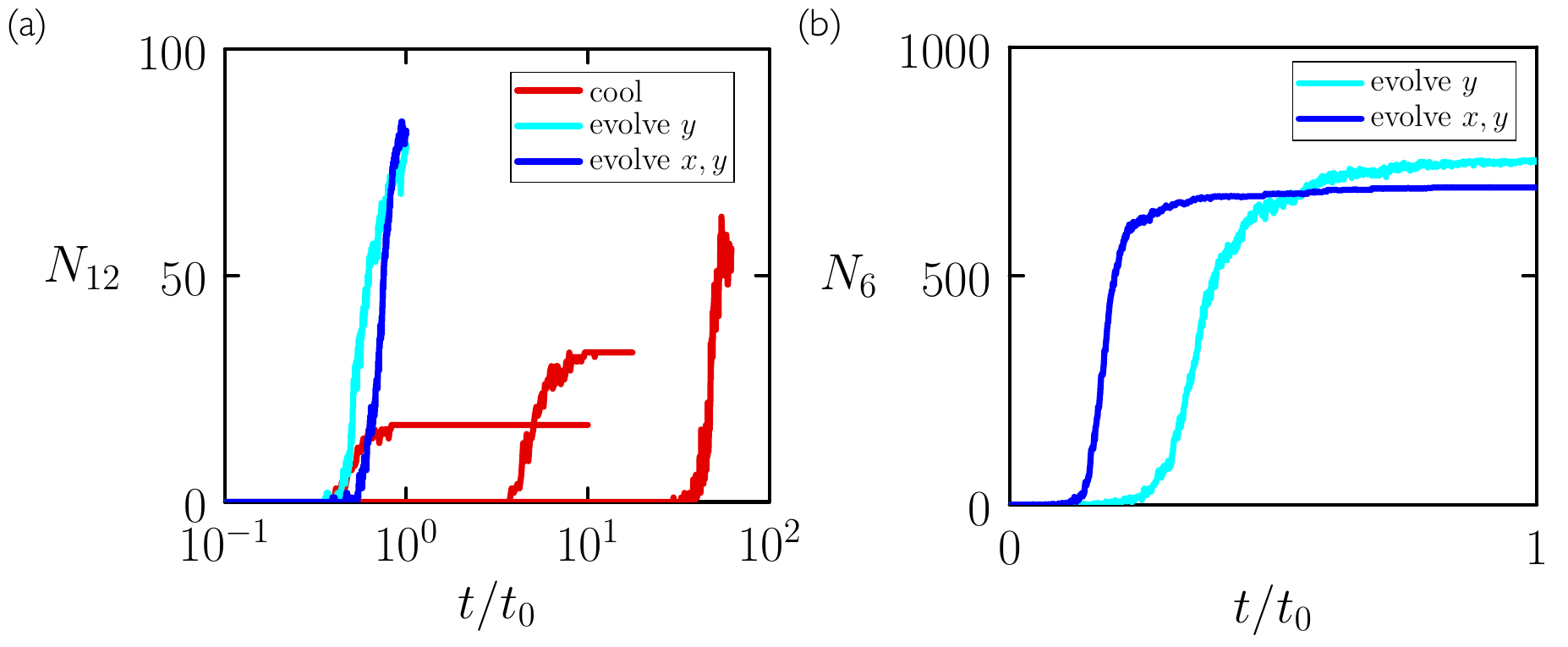} 
   \caption{\label{fig_yield}(a) Yield of the self-assembly process using the best generation-27 genome (particle and potential) of \f{fig_12_gons} (blue line) compared to the yield obtained using a human-designed particle\c{whitelam2016minimal} combined with evolutionary learning of the protocol (light blue). Also shown are three simulations (red) using a human-designed particle and protocols, the latter consisting of slow cooling at three rates at fixed chemical potential. (b) Yield of the self-assembly process using the best generation-18 genome from \f{fig_collection}(c) and \f{fig_6_gons} (blue line), compared to that using a particle with perfect three-fold rotational symmetry combined with evolutionary learning of the protocol (light blue).}
\end{figure*}
\clearpage

\begin{figure*}[] 
   \centering
\includegraphics[width=\linewidth]{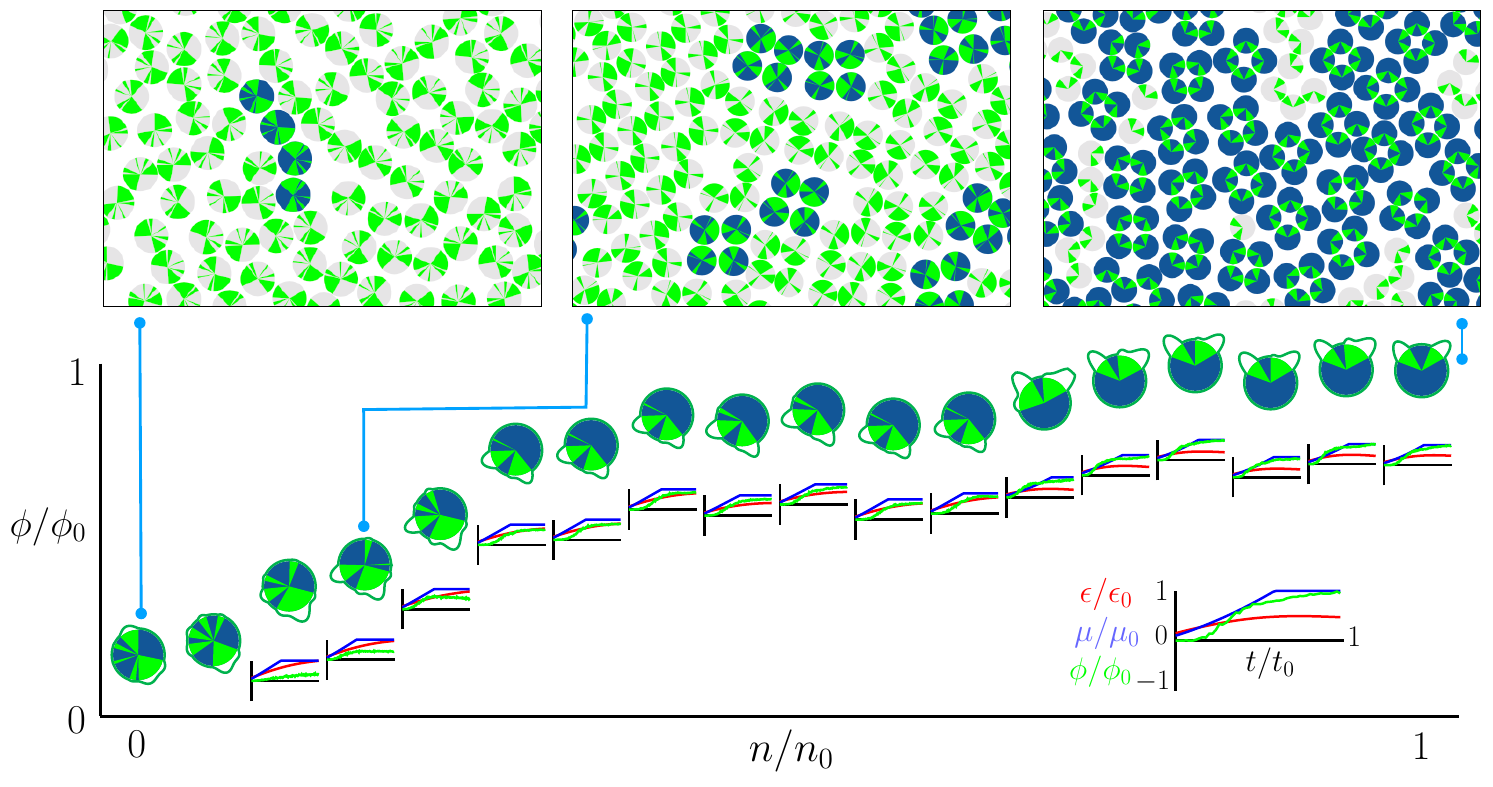} 
   \caption{\label{fig_4_mers} Evolutionary learning with $\phi=C_4$, the number of 4-mers, interacting clusters of 4 particles (shown darker in images). The format of the figure is the same as \f{fig_12_gons}. Parameters: $\phi_0=350$ 4-mers, $\epsilon_0=\mu_0=20\, \kt$, $n_0=17$ generations, $t_0=10^9$ Monte Carlo steps.}
\end{figure*}

\clearpage
\begin{figure*}[] 
   \centering
\includegraphics[width=\linewidth]{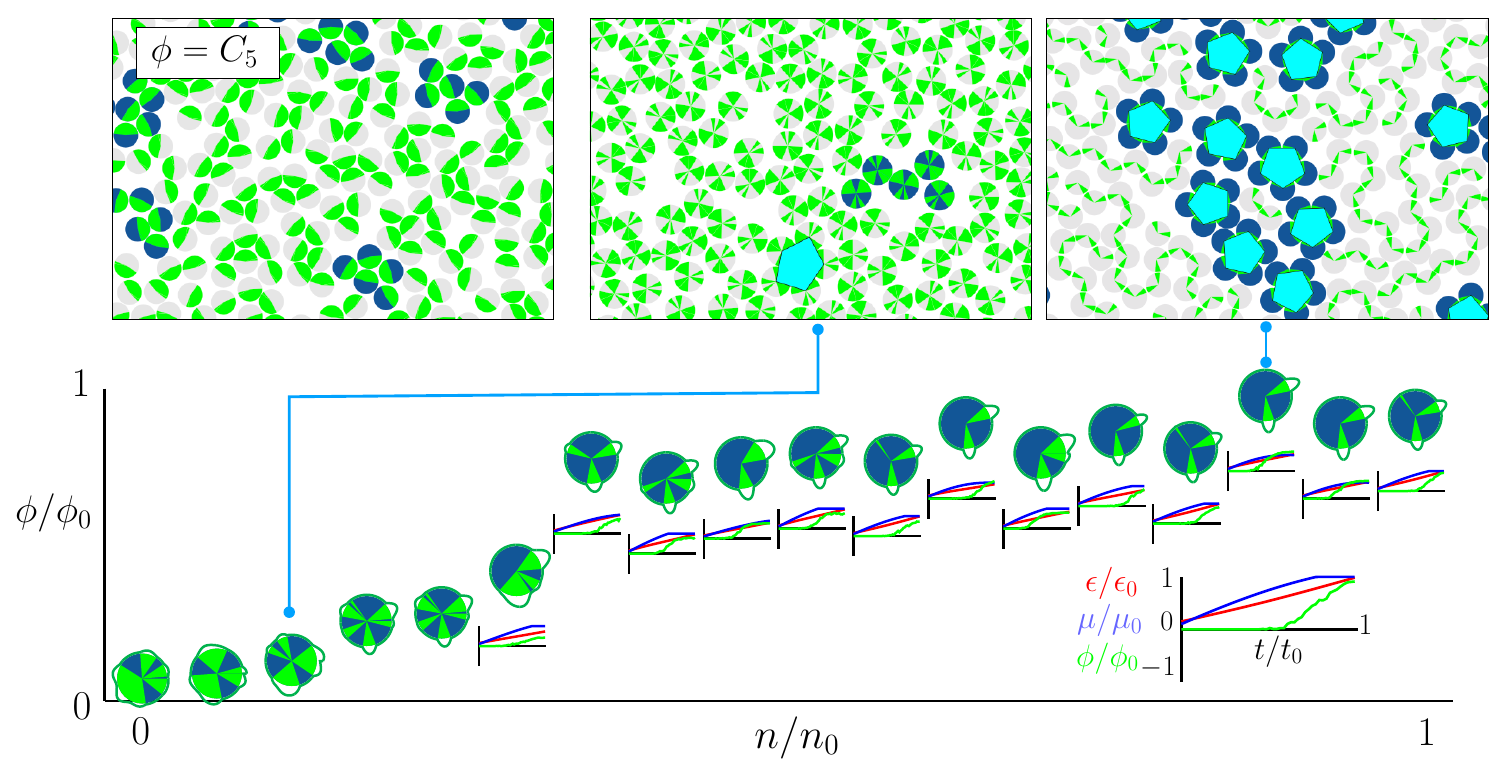} 
   \caption{\label{fig_5_mers_gons} Evolutionary learning with $\phi=\min(C_5,N_5)$, the number of 5-mers (shown darker in images) and 5-gons (shown light blue). The format of the figure is the same as \f{fig_12_gons}, with the exception of the top-left image: that shows the outcome of learning after 17 generations with $\phi=C_5$. Evolution of pentagonal clusters requires the dual goal of 5-mers and 5-gons. Parameters: $\phi_0=125$ 5-mers \& 5-gons, $\epsilon_0=\mu_0=20\, \kt$, $n_0=17$ generations, $t_0=10^9$ Monte Carlo steps.}
\end{figure*}

\clearpage

\begin{figure*}[] 
   \centering
\includegraphics[width=\linewidth]{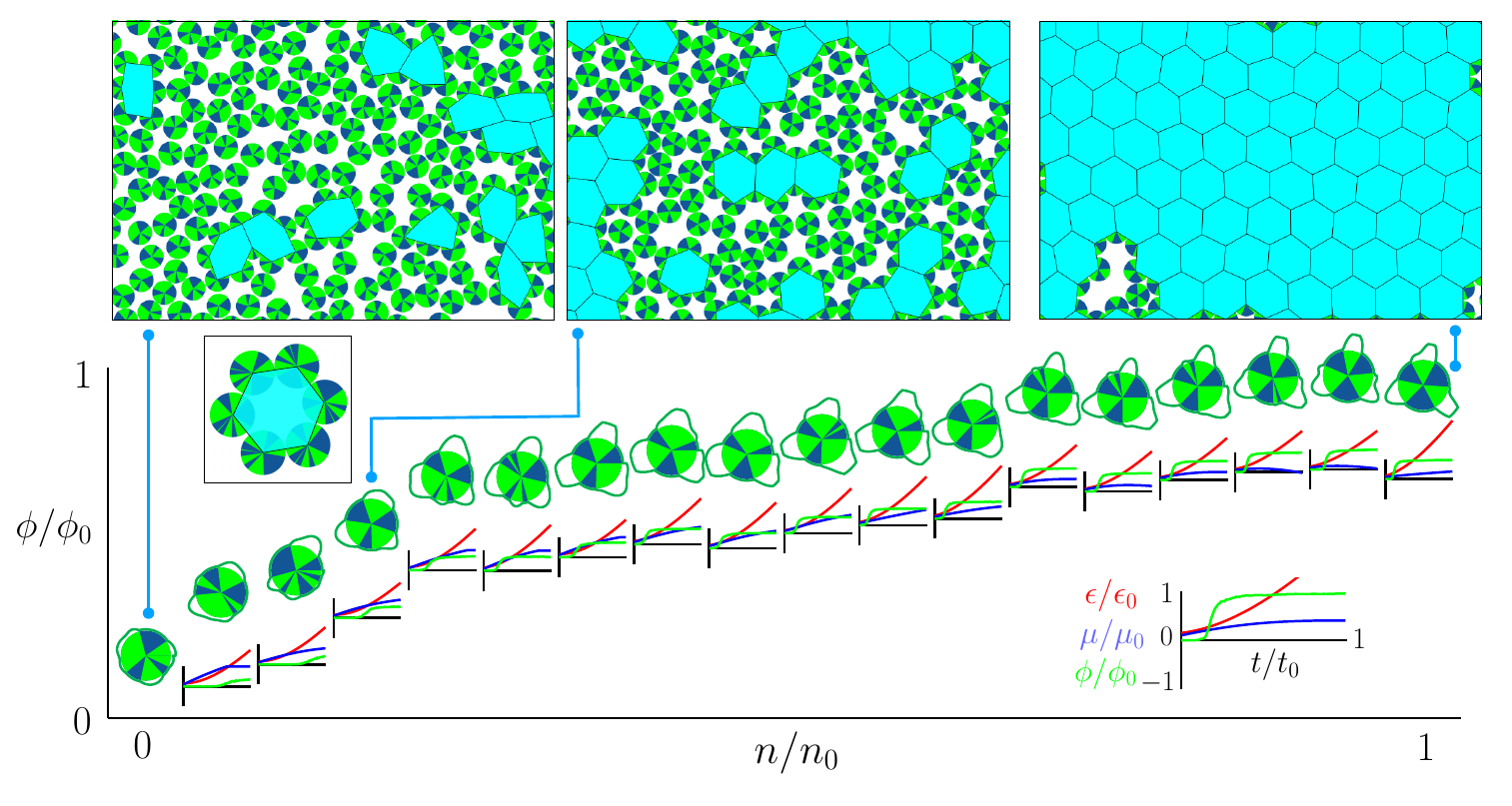} 
   \caption{\label{fig_6_gons} Evolutionary learning with $\phi=N_6$, the number of 6-gons, convex pores of 6 interacting particles (shown light blue: see boxed image). The format of the figure is the same as \f{fig_12_gons}. Parameters: $\phi_0=700$ 6-gons, $\epsilon_0=\mu_0=20\, \kt$, $n_0=17$ generations, $t_0=10^9$ Monte Carlo steps.}
\end{figure*}


\clearpage

\begin{figure}[] 
   \centering
\includegraphics[width=\linewidth]{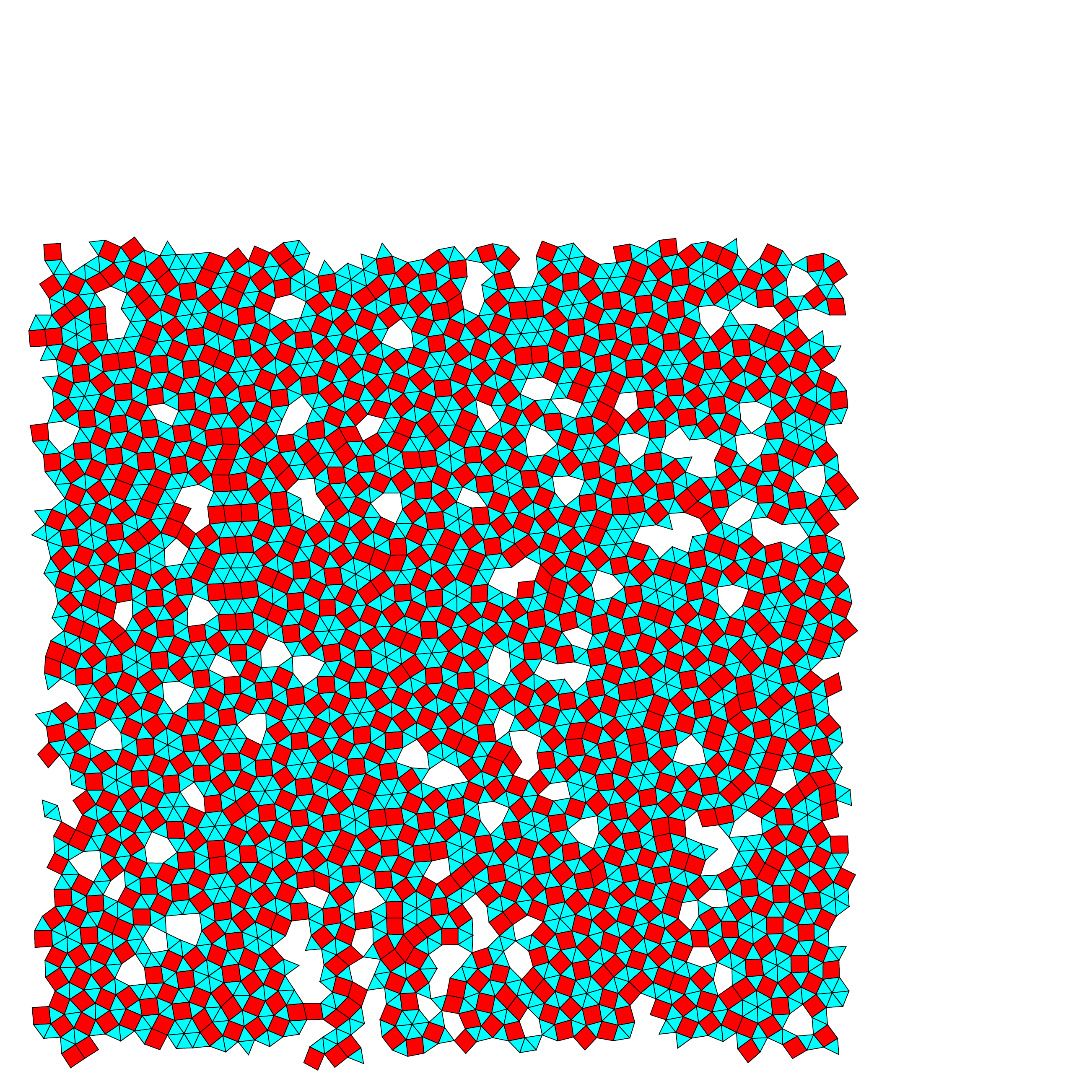} 
   \caption{\label{fig_novelty2} One of the materials identified by the novelty search of \f{fig_novelty} shows the $\sigma$, H, and Z binding motifs characteristic of dodecagonal quasicrystals\c{van2012formation}. To produce this figure we used the particle- and protocol design identified by the learning algorithm of \f{fig_novelty}, and ran a trajectory of length $20 t_0$. The particles underlying the pattern of 3-gons and 4-gons are not shown.}
\end{figure}

\end{document}